\documentclass[11pt]{article}
\usepackage[a4paper,bindingoffset=0.2in,%
            left=0.9in,right=0.9in,top=1in,bottom=1in,%
            footskip=.25in]{geometry}

\usepackage{units}        
\usepackage{gensymb}      
\usepackage{textcomp}     
\usepackage{graphicx}     
\usepackage[labelfont=bf,font={small,sf}]{caption}
\usepackage{subcaption}
\usepackage{float}        
\usepackage{pgfplots}
\usepackage{pgfplotstable}
\usepackage{tikzscale}
\usepackage{etoolbox}

\usepackage[boldvectors,boldmatrices,bladraft]{blatex}
\usepackage{csquotes}
\usepackage{silence}
\usepackage[style=ieee]{biblatex}
\usepackage{enumitem}

\def\Linkcolor{blue}

\hypersetup{
  final,
  linktocpage=true,
  colorlinks=true,
  linkcolor=\Linkcolor,
  citecolor=\Linkcolor,
  filecolor=\Linkcolor,
  urlcolor=\Linkcolor
}

\NewDocumentCommand{\wraptikz}{s +m}{%
  \IfBooleanTF{#1}{%
    \tikzexternaldisable%
    #2
    \tikzexternalenable%
  }{
    #2
  }
}

\NewDocumentCommand{\eatme}{s +m}{%
  \IfBooleanTF{#1}{%
    #2%
  }{}%
}

\newcommand*{\dimx}{d}
\newcommand*{\dimxtot}{D}
\newcommand*{\dimy}{m}
\newcommand*{\dimytot}{M}
\newcommand*{\dims}{s}
\newcommand*{\dimg}{g}
\newcommand*{\numO}{L}
\newcommand*{\numG}{G}

\newcommand*{\xorg}{{\mathring{\vmx}}}

\newcommand*{\Tset}{\mathcal{T}}

\newcommand*{\partI}{\mathcal{I}}
\newcommand*{\partilde}{\widetilde{\partI}}
\newcommand*{\net}{\mathcal{N}}

\NewDocumentCommand{\SparseVecs}{s o}{%
  \IfBooleanTF{#1}{\widetilde{\Sigma}}{\Sigma}_{%
    \IfValueTF{#2}{#2}{\dims}%
  }%
}
\NewDocumentCommand{\eSparseVecs}{s o}{%
  \IfBooleanTF{#1}{\widetilde{\mathcal{E}}}{\mathcal{E}}_{%
    \IfValueTF{#2}{#2}{\dims}%
  }%
}
\DeclareDocumentCommand{\BSparseVecs}{s d<> o}{%
  \IfBooleanTF{#1}{%
    \SparseVecs*[%
      \IfValueTF{#2}{#2}{\partI},%
      \IfValueTF{#3}{#3}{\dims}%
    ]%
  }{%
    \SparseVecs[%
      \IfValueTF{#2}{#2}{\partI},%
      \IfValueTF{#3}{#3}{\dims}%
    ]%
  }
}
\DeclareDocumentCommand{\eBSparseVecs}{s d<> o}{%
  \IfBooleanTF{#1}{%
    \eSparseVecs*[%
      \IfValueTF{#2}{#2}{\partI},%
      \IfValueTF{#3}{#3}{\dims}%
    ]%
  }{%
    \eSparseVecs[%
      \IfValueTF{#2}{#2}{\partI},%
      \IfValueTF{#3}{#3}{\dims}%
    ]%
  }
}

\DeclareDocumentCommand{\bnorm}{s o m m}{%
  \IfBooleanTF{#1}{\norm*{#3}}{\norm{#3}}_{\IfValueTF{#2}{#2}{\partI},#4}%
}

\newcommand*{\bsupp}{\supp_\partI}

\NewDocumentCommand{\includepdf}{m}{%
  \noindent\includegraphics[width=\textwidth]{#1}%
}

\NewDocumentCommand{\ConjSym}{o}{%
  \mathbb{X}_{\IfValueTF{#1}{#1}{\dimx}}%
}

\newcommand*{\fb}{f_\mathrm{b}}
\NewDocumentCommand{\Bandlimited}{o}{%
  \IfValueTF{#1}{%
    B_\fourier([-#1,#1])%
  }{%
    B_\fourier([-\fb,\fb])%
  }%
}

\newcommand*{\zhat}{\hat{\vmz}}

\newcommand*{\Uset}{\mathcal{U}}

\newcommand*{\RecGeneric}{\widetilde{\Delta}}

\NewDocumentCommand{\RecCP}{o}{%
  \RecGeneric_{\IfValueTF{#1}{#1}{\partI}}^\mathrm{PV}%
}
\NewDocumentCommand{\RecCor}{o}{%
  \RecGeneric_{\IfValueTF{#1}{#1}{\partI}}^\mathrm{corr}%
}
\NewDocumentCommand{\RecHT}{o}{%
  \RecGeneric_{\IfValueTF{#1}{#1}{\partI}}^\mathrm{ht}%
}

\NewDocumentCommand{\RecNormNcCP}{o}{%
  \Delta_{\IfValueTF{#1}{#1}{\partI}}^\mathrm{nc}%
}
\NewDocumentCommand{\RecNormCP}{o}{%
  \Delta_{\IfValueTF{#1}{#1}{\partI}}^\mathrm{PV}%
}
\NewDocumentCommand{\RecNormReg}{o}{%
  \Delta_{\IfValueTF{#1}{#1}{\partI}}^\proj{}%
}
\NewDocumentCommand{\RecNormCor}{o}{%
  \Delta_{\IfValueTF{#1}{#1}{\partI}}^\mathrm{corr}%
}
\NewDocumentCommand{\RecNormHT}{o}{%
  \Delta_{\IfValueTF{#1}{#1}{\partI}}^\mathrm{ht}%
}
\NewDocumentCommand{\RecNormRegHT}{o}{%
  \Delta_{\IfValueTF{#1}{#1}{\partI}}^{\proj{}\text{-}\mathrm{ht}}%
}

\DeclareDocumentCommand{\suppnorm}{s o m}{%
  \IfBooleanTF{#1}{\norm*{#3}}{\norm{#3}}_{(\IfValueTF{#2}{#2}{\dims})}%
}

\def\sigset{\Sigma_{\partI,\dims}}
\def\sigsetO{\Omega}
\def\radFrob{\rho_\mathrm{F}}
\def\radOp{\rho_{2\to2}}
\def\gamFunc{\gamma_2}
\def\opimg{\mathcal{M}}
\def\orthoconst{\omega_\partI}
\def\Vhat{{\widehat{V}}}
\def\Vtilde{{\widetilde{V}}}
\def\opimghat{{\widehat{\opimg}}}

\NewDocumentCommand{\coherence}{o}{\mu_{\IfValueTF{#1}{#1}{\partI}}}

\newcommand*{\ktermerror}[1]{\sigma_{#1}}

\pgfplotsset{compat=newest}
\usepgfplotslibrary{external,colormaps}
\usetikzlibrary{plotmarks,arrows.meta,calc}
\newlength\figureheight
\newlength\figurewidth
\pgfplotsset{
  every axis/.append style={
    filter discard warning=false,
    label style={font=\Large},
    tick label style={font=\normalsize},
    legend style={font=\normalsize, inner sep=4pt}
  }
}

\usepackage{xcolor}

\graphicspath{ {img/} }

\begin{document}

\title{The Restricted Isometry Property of Block Diagonal Matrices for
Group-Sparse Signal Recovery}
\author{Niklas Koep and Arash Behboodi and Rudolf Mathar\thanks{
Institute for Theoretical Information Technology, RWTH  Aachen  University, Germany }}
\date{}

\maketitle

\Env{abstract}{
  Group-sparsity is a common low-complexity signal model with widespread
  application across various domains of science and engineering. The recovery
  of such signal ensembles from compressive measurements has been extensively
  studied in the literature under the assumption that measurement operators are
  modeled as densely populated random matrices. In this paper, we turn our
  attention to an acquisition model intended to ease the energy consumption of
  sensing devices by splitting the measurements up into distinct signal blocks.
  More precisely, we present uniform guarantees for group-sparse signal
  recovery in the scenario where a number of sensors obtain independent partial
  signal observations modeled by block diagonal measurement matrices. We
  establish a group-sparse variant of the classical \acf{RIP} for block
  diagonal sensing matrices acting on group-sparse vectors, and provide
  conditions under which subgaussian block diagonal random matrices satisfy
  this \acs{GRIP} with high probability. Two different scenarios are considered
  in particular. In the first scenario, we assume that each sensor is equipped
  with an independently drawn measurement matrix. We later lift this
  requirement by considering measurement matrices with constant block diagonal
  entries. In other words, every sensor is equipped with a copy of the same
  prototype matrix. The problem of establishing the \acs{GRIP} is cast
  into a form in which one needs to establish the concentration behavior of the
  suprema of chaos processes which involves estimating Talagrand's $\gamFunc$
  functional. As a side effect of the proof, we present an extension to
  Maurey's empirical method to provide new bounds on the covering number of
  sets consisting of finite convex combinations of possibly infinite sets.
}

\section{Introduction}

A common problem in modern signal processing applications is that of sampling
signals containing only a limited amount of information imposed by some type of
low-complexity structure. The most common low-complexity structure by far
manifests in the form of signal sparsity in a suitable basis or more generally
in an overcomplete dictionary \cite{rauhut2008compressed} or frame
\cite{casazza2012finite}. The field of \acl{CS} was founded on the very idea
that the number of samples required to acquire, and represent such signals
should be on the order of the information-theoretic rather than the
linear-algebraic dimension of the ambient signal space. This was the result of
a series of landmark papers due to Cand\`es, Tao, Romberg
\cite{Candes2005,Candes2006,Candes2006-2,Cands2006NearOptimalSR} and Donoho
\cite{Donoho2006} who first showed that every $\dimx$-dimensional vector $\vmx$
containing at most $\dims$ nonzero coefficients can be perfectly reconstructed
from $\Omega{\dims\log(\dimx/\dims)}$ nonadaptive measurements of the form $\vmy
= \vmA\vmx$ with $\vmA \in \C^{\dimy \times \dimx}$, and $\dimy \ll \dimx$,
assuming that the measurement matrix $\vmA$ satisfies certain structural
conditions. While the deterministic construction of such matrices with
provably optimal scaling in terms of the information dimension of signals
remains a yet unsolved problem, it is by now a well-established fact that a
multitude of random ensembles in the class of subgaussian random variables are
able to capture just enough information about signals of interest to allow for
them to be reconstructed in polynomial time by a variety of different
algorithms. More recently, it was also demonstrated that similar results can be
obtained for more heavy-tailed ensembles such as measurement matrices populated
by independent copies of subexponential random variables
\cite{Dirksen2018OnTG}. Moreover, it was established very early on that
measurement matrices constructed from randomly chosen samples of basis
functions in \acp{BOS} could provide similar guarantees as unstructured
ensembles. Typical examples of this class of structured random matrices are
partial Fourier transform matrices, partial circulant matrices generated by a
subgaussian random vector or subsampled Hadamard matrices.

While unstructured random matrices are highly desirable from a theoretical
perspective, practitioners are not usually free to choose measurement matrices
at a whim. Instead, in most engineering applications, most structural
properties of the measurement system are predetermined by the application at
hand. In this work, we consider another class of structured random matrices at
the intersection of purely random and highly structured measurement ensembles.
In particular, we consider block diagonal measurement matrices whose blocks are
either independent or identical copies of a dense subgaussian random matrix.
Such measurement models appear in various applications of interest like
\ac{DCS} \cite{Duarte2005DistributedCS}, and the so-called \ac{MMV} model in
which one obtains multiple independent snapshots of a signal whose
low-complexity structure is assumed to be stationary in time
\cite{Rao2004SparseST,Chen2006TheoreticalRO}. Moreover, such block-wise
measurement paradigms have previously been studied in image acquisition systems
in order to ease both storage and energy demands of sensors, and recovery
algorithms \cite{Gan2007BlockCS}.
 This acquisition model was previously addressed by Eftekhari \etal in
\cite{Eftekhari2012TheRI} where the authors establish a lower bound on the
number of measurements for subgaussian block diagonal matrices to satisfy the
classical \ac{RIP}, implying stability and robustness guarantees for recovery
of sparse vectors. More recently the model was employed by Palzer and Maly
in the context of quantized \ac{DCS} with 1-bit observations.

In the present
work, we extend the results of \cite{Eftekhari2012TheRI} to more structured
signal sets, namely those whose nonzero coefficients appear in groups. {There are many works in compressed sensing literature on different signal structures. A very general notion of structure is discussed in \cite{chandrasekaran_convex_2012}, where the signal structure is captured through the notion of atomic sets. Atomic sets contain finite or infinite number of atoms. The signal set is constructed by linear combinations of finite subset of atoms. A similar notion appears in \cite{plan_generalized_2016} where the  structure is given directly by specifying the signal set. The set can potentially have lower complexity measured by its Gaussian width. Despite the generality of these frameworks, the signal recovery in these cases involves minimizing a computationally complex norm. {For example, for atomic sets, the recovery is done
via minimizing the atomic norm, which is defined in terms of convex hull of the respective atomic set. In general, this convex hull does not have a tractable representation, and even testing membership in the set is in general undecidable. When the considered atomic set is an algebraic variety, the atomic norm minimization problem can be solved approximately by semidefinite programming. See  \cite[Section 4]{chandrasekaran_convex_2012} for detailed discussions on computing atomic norms.} 
There are other notions of structure for which the signal recovery can be still done by minimizing simpler norms, say combination of the $\ell_p$-norms for different $p\geq 1$. Group sparsity is an example of structured sparsity for which the corresponding norm remains comparable to the $\ell_1$ norm. } This type of structured sparsity frequently arises in audio \cite{Adalbjornsson2013}
and image signal processing \cite{Wright2009}, \eg, in modeling the transform
coefficients of the wavelet packet transform \cite{baraniuk2010model}. Other
common applications include multi-band reconstruction and spectrum sensing
\cite{Mishali2007, Polo2009}, sparse subspace clustering \cite{Elhamifar2009},
as well as measurement of gene expression levels \cite{Parvaresh2008} and
protein mass spectroscopy \cite{Tibshirani2005}.

{This setup is particularly relevant for sensing scenarios where the goal is to reconstruct a signal based on multiple independent linear acquisitions. Notable examples are medical imaging and electron microscopy problems. Multiple linear measurements $\boldsymbol{\Phi}_i$, $1\leq i\leq L$, are used to obtain the image of an entity. The image  changes slightly from one measurement to another, therefore giving rise to $L$ different inputs in each measurement. However, the main features remain stable across measurements. One way to capture this stability is to consider a shared sparsity structure and  support between images in a proper sparsity basis matrix $\boldsymbol{\Psi}$. The whole acquisition problem can be then modeled as a single linear inverse problem. The sensing matrix is a block diagonal matrix $\mathbf{A}$ with blocks as $\boldsymbol{\Phi}_i$. The input signal is modeled as $\boldsymbol{\Psi}\mathbf{x}$ where $\mathbf{x}$ is a group sparse signal. In this paper, we consider such a scenario in its full generality with arbitrary number of groups and group structures.  }

\subsection{Notation}

Throughout the paper, we denote matrices by uppercase boldface letters, vectors
by lowercase boldface letters, and scalars by regular type symbols. For an
integer $n \in \N$, we use the common shorthand notation $[n] := \set{1,
\ldots, n} = [1, n] \cap \N$. Given a norm $\norm{\cdot}_\theta$ on $\C^n$
depending on some abstract parameter set $\theta$, we write $\ball{\theta}^n$
for the norm ball associated with $\norm{\cdot}_\theta$, \ie, $\ball{\theta}^n
= \set{\vmx \in \C^n}[\norm{\vmx}_\theta \leq 1]$. Even though we mostly work
in $\C^n$, we denote by $\function{\inner{\cdot}{\cdot}}{\C^n}{\C^n}$ the
bilinear form defined by $\inner{\vma}{\vmb} = \sum_{i=1}^n a_i b_i$ for
$\vma,\vmb \in \C^n$ rather than a sesquilinear form inducing an inner product
on $\C^n$. As such, the canonical $\ell_2$-norm on $\C^n$ is induced by
$\norm{\vma}_2 = \inner{\vma}{\conj{\vma}} = \inner{\conj{\vma}}{\vma}$ where
$\conj{\vma}$ denotes the complex conjugate of $\vma \in \C^n$. Finally, we
denote the unit Euclidean sphere in $\C^n$ by $\sphere{n-1}$. To ease notation,
we will make frequent use of the following asymptotic notation: given two
scalars $a,b \in \R$, we write $a \lesssim b$ if there exists a universal
constant $C > 0$ such that $a \leq C b$ holds. Similarly, we write $a \gtrsim
b$ to mean $a \geq C b$.

\subsection{Summary of Contributions}

The contributions of this paper are as follows. We establish the so-called
\acl{GRIP} for subgaussian block diagonal matrices acting on group-sparse
vectors. We consider two distinct variations of block diagonal measurement
matrices. First, we assume that each block of a measurement matrix is an
independent copy of a subgaussian random matrix. In the second case, it is
assumed that the block diagonal sensing matrix has constant block diagonal,
\ie, each block is a copy of one prototype matrix drawn at random from a
subgaussian distribution. Appealing to this \acl{GRIP}, it is shown that
group-sparse vectors can be stably and robustly reconstructed from partial
observations obtained via block diagonal measurement operators. The scaling
behavior obtained for such matrices to satisfy the \acs{GRIP} matches up to
logarithmic factors the lower bound on the number of required measurements for
suitably chosen unitary bases. Furthermore, we show that our results reduce to
previous results reported in \cite{Eftekhari2012TheRI}. Motivated by ideas in
\opcit, we relate the problem of establishing the \acs{GRIP} to estimating
certain geometric quantities associated with the suprema of chaos processes
involving Talagrand's $\gamFunc$-functional. Since the methods employed in
\cite{Eftekhari2012TheRI} do not directly apply to the group-sparse setting, we
propose an alternative method to estimate the covering number at higher scales.
In particular, we extend Maurey's empirical method to sets which do not admit a
polytope representation. As a side effect of the proof, we therefore provide a
generalization of Maurey's lemma to provide new bounds on the covering number
of sets that consist of finite convex combinations of possibly infinite sets.

\subsection{Organization}

The paper is organized as follows. The definition of group-sparse vectors and
the underlying sensing model is stated in \blaref{sec:model}. The main results
are presented in \blaref{sec:mainresults} where we also discuss connections to
other related results in the literature. The proofs of our main theorems are
given in \blaref{sec:proof_bdgrip} and \blaref*{sec:shared_sensing_matrix}.
Finally, we conclude the paper in \blaref{sec:conclusion}.

\section{Group Sparse Signals and Distributed Sensing}
\label{sec:model}

We consider the problem of recovering signals with a low-complexity structure
in the form of group-sparsity from partial observations. These partial
observations are modeled by means of block diagonal measurement matrices. In
particular, we assume a vector $\vmx \in \C^\dimxtot$, which we decompose into
$\numG$ nonoverlapping groups, is observed by $\numO$ sensors.
{In general, each sensor $i$, $i\in[\numO]$ has a $d_i$ dimensional observation such that $D=\sum_{i=1}^\numO d_i$.
For simplicity of presentation,
we assume that $\dimxtot$ is an integer multiple of $\numO$ such that $\dimxtot
= \dimx\numO$ with $\dimx \in \N$ and $d_i=\dimx$ for all $i\in[\numO]$.
All the arguments in this paper can be reworked for the former case.} To define the group-sparsity structure on
$\vmx$, we partition the set $[\dimxtot]$ into $\numG$ groups as follows.

\Definition[Group partition]{\label{def:group_partition}
  A collection $\partI = \set{\partI_1,\ldots,\partI_\numG}$ of subsets $\partI_i
  \subseteq [\dimxtot] := \set{1,\ldots,\dimxtot}$ is called a \textit{group
  partition} of $[\dimxtot]$ if $\partI_i \cap \partI_j = \emptyset$ for all $i
  \neq j$, and $\bigcup_{i=1}^\numG \partI_i = [\dimxtot]$.
}

Note that this definition does not assume that the elements in $\partI_i$ are
consecutive indices, nor that the cardinality of the individual sets are
identical. For simplicity of notation, we denote the size of each group by
$\dimg_i := \card{\partI_i}$ such that $\sum_{i=1}^G \dimg_i=D$. Moreover, we
denote the cardinality of the biggest group by $\dimg := \max_{i \in
[\numG]}\dimg_i$. We emphasize that we only consider nonoverlapping group
partitions in contrast with other works which often allow for coefficient
groups to overlap, rendering it a nontrivial task to decompose a given vector
$\vmx$ into individual groups. Some authors refer to this variant as strict
group-sparsity. As we will discuss in \blaref{sec:connection_to_dcs}, the
flexibility in the group structure leads to certain adversarial examples which
will not allow us to correctly estimate the number of measurements required to
stably and robustly recover certain signals.

To properly define the concept of group-sparsity, we introduce the following
notation. Denote by $\vmx_{\partI_i} \in \C^{\dimxtot}$ the restriction of
$\vmx$ to the indices in $\partI_i$, \ie, $(\vmx_{\partI_i})_j = x_j \cdot
\binarize{j \in \partI_i}$ for $j \in [\dimxtot]$ where $\binarize{j \in
\partI_i}$ denotes the binary indicator function of the event $\set{j \in
\partI_i}$. Then a signal $\vmx$ is called $\dims$-group-sparse (\wrt the group
partition $\partI$) if it is supported on at most $\dims$ groups, \ie, $\vmx =
\sum_{i \in S} \vmx_{\partI_i}$ for some $S \subset [\numG]$ with $\card{S} \leq
\dims$. We also define the following family of mixed norms.

\Definition[Group $\ell_{\partI,p}$-norms]{\label{def:group_lp_norms}
  Let $\vmx \in \C^\dimxtot$. Then, for $p \geq 1$, the group
  $\ell_{\partI,p}$-norm on $\C^\dimxtot$ is defined as
  \Equation*{
    \bnorm{\vmx}{p} := \parens{\sum_{i=1}^\numG
    \norm{\vmx_{\partI_i}}_2^p}^{1/p}.
  }
}

As is customary in the literature on sparse recovery, we extend the notation
$\bnorm{\cdot}{p}$ to $p=0$ in which case $\norm{\cdot}_{\partI,0}$ corresponds
to the group $\ell_{\partI,0}$-pseudonorm which counts the number of groups a
vector is supported on:
\Equation*{
  \bnorm{\vmx}{0} := \card{\set{i \in [\numG]}[\vmx_{\partI_i} \neq \vmzero]}.
}
With this definition in place, we define the set
\Equation*{
  \sigset = \set{\vmx \in \C^\dimxtot}[\norm{\vmx}_{\partI,0} \leq \dims]
}
of $\dims$-group-sparse vectors \wrt the group partition $\partI$. In most
practical real-world settings, it is unlikely that signals of interest
precisely adhere to this stringent signal model. Instead, one usually
assumes that real-world signals are only well-approximated by elements of
$\sigset$. In particular, with the definition of the best $\dims$-term group
approximation error
\Equation*{
  \ktermerror{\dims}(\vmx)_{\partI,1} = \inf_{\vmz \in \sigset} \bnorm{\vmx -
  \vmz}{1},
}
one commonly considers so-called \textit{compressible vectors} which are
characterized by the fact that $\ktermerror{\dims}(\cdot)_{\partI,1}$ rapidly
decays as $\dims$ increases.

As hinted at before, we consider a measurement setup in which we observe an
$\dims$-group-sparse or compressible signal $\vmx$ by means of a block diagonal
matrix $\vmA$ consisting of $\numO$ blocks, namely
\Equation*{
\vmA=\PMatrix{
    \vmPhi_1 & & \\
    & \ddots & \\
    & & \vmPhi_{\numO}
  }.
}
 However, we assume that we only have
access to the signal $\vmx \in \sigset$ in terms of its basis expansion $\vmz$
in a unitary basis $\vmPsi \in \Ugroup(\dimxtot) := \set{\vmU \in \C^{\dimxtot
\times \dimxtot}}[\adj{\vmU}\vmU = \Id_\dimxtot]$. The measurement model
therefore reads
\Equation{\label{eqn:measurements}
  \vmy = \diag{\vmPhi_l}_{l=1}^\numO \vmz = \diag{\vmPhi_l}_{l=1}^\numO \vmPsi
  \vmx = \vmA\vmPsi\vmx.
}
We will also consider an alternative measurement model in which each sensor is
equipped with a copy of the same matrix $\vmPhi$, \ie, $\vmPhi_l = \vmPhi
$ for all $l \in [\numO]$.
Ultimately, our goal in this paper is to provide a sufficient condition for
stable and robust recovery of group-sparse signals by establishing a suitable
\ac{RIP} property of block diagonal matrices acting on group-sparse vectors.

\subsection{{Subgaussian  Random Variables and Vectors}}

{
We first recall some definitions used in this paper.
\Definition[Subgaussian random variable]{
  A zero mean random variable $X$ is called \emph{subgaussian} if there exists
  a constant $C>0$ such that
  \Equation*{
    \E(\exp(X^2/C^2))\leq 2.
  }
  The subgaussian norm of $X$, also known as Orlicz norm of $X$, is defined by
  \Equation*{
    \subgnorm{X}=\inf\set{C>0}[\E(\exp(X^2/C^2))\leq 2].
  }
}
The definition can be extended to subgaussian random vectors by considering its one dimensional projections.
\Definition[Subgaussian random vector]{
A random vector $\vmX$ in $\R^\dimx$ is called \emph{subgaussian} if the one dimensional projections $\inner{\vmX}{\vmx}$ 
are subgaussian random variables for all $\vmx\in\R^\dimx$. The subgaussian norm of $\vmX$ is defined by
  \Equation*{
    \subgnorm{\vmX}=\sup_{\vmx\in S^{\dimx-1}} \subgnorm{\inner{\vmX}{\vmx}}.
  }
}
A random vector $\vmX$ in $\R^\dimx$ is called isotropic if $\E(\vmX\vmX^T)=\vmI_\dimx$ with $\vmI_\dimx$ the identity matrix. The notation $a \gtrsim_\tau b$ ($a\lesssim_\tau a$) indicates that there is a function $C(\tau)$ such that $a \geq C(\tau) b$ ($a\leq C(\tau) b$). 
}

\section{Signal Recovery with Block Diagonal
\texorpdfstring{\Acs{GRIP}}{Group-RIP} Matrices}
\label{sec:mainresults}

The analysis of both sensing paradigms introduced in the previous section
relies on the so-called \ac{GRIP}---a generalization of the well-known
\acl{RIP} modeled on the block-sparse \ac{RIP} first introduced in
\cite{Eldar2009RobustRO}.


\Definition[\Acl{GRIP}]{\label{def:group_rip}
{A matrix $\vmA \in \C^{\dimytot \times \dimxtot}$ is said
  to satisfy the \acfi{GRIP} of order $\dims$ if, for $\delta \in (0,1)$,
  \Equation{\label{eqn:group_rip}
    (1-\delta)\norm{\vmx}_2^2 \leq \norm{\vmA\vmx}_2^2 \leq
    (1+\delta)\norm{\vmx}_2^2 \quad\Forall \vmx \in \sigset.
  }
  The smallest constant $\delta_\dims \leq \delta$ for which
  \blaref{eqn:group_rip} holds is called the \acfi{GRIC} of $\vmA$.}
}

In combination with the above definition, the result due to Gao and Ma
established in \cite{Gao2017ANB} which we will introduce next then implies
stable and robust recovery of group-sparse signals {by solving the quadratic constrained basis pursuit problem given in \blaref{prob:group_qcbp} }. While the signal model
employed in \cite{Gao2017ANB} assumes that the indices in each group $\partI_i$
are linearly increasing, \ie, the signals are assumed to be block- rather than
group-sparse, the proof of \thmName~1 in \cite{Gao2017ANB} does not explicitly
rely on this structure. Furthermore, the result was originally proven in the
real setting, but the proof is easily extended to the complex case. Note that
such a stability and robustness result was already established in the seminal
work of Eldar and Mishali \cite{Eldar2009RobustRO}, albeit with the necessary
condition $\delta_{2\dims} < \sqrt{2} - 1$ on the \acs{GRIP} constant. These
results therefore also extend to more general group partitions as defined in
\blaref{def:group_partition}. The precise statement of this generalization is
stated in the following result. For the sake of being self-contained, we
provide a proof in \blaref{app:proof_thm_group_rip}.

\Theorem{\label{thm:group_rip_guarantee}
  Let $\tilde{\vmA} \in \C^{\dimytot \times \dimxtot}$ be a matrix satisfying
  the \acl{GRIP} of order $2\dims$ with constant $\delta_{2\dims} <
  4/\sqrt{41}$. Then for any $\xorg \in \C^\dimxtot$, and $\vmy =
  \tilde{\vmA}\xorg + \vme$ with $\norm{\vme}_2 \leq \epsilon$, {a} solution
  $\opt{\vmx}$ of the program
  \Equation{\label{prob:group_qcbp}
    \tag{P$_{\partI,1}$}
    \MinProblem{\bnorm{\vmx}{1}}{\norm{\tilde{\vmA}\vmx - \vmy}_2 \leq
    \epsilon}
  }
  satisfies
  \Equation*{
    \norm{\xorg - \opt{\vmx}}_2 \leq C
    \frac{\ktermerror{\dims}(\xorg)_{\partI,1}}{\sqrt{\dims}} + D \epsilon
  }
  where the constants $C,D > 0$ only depend on $\delta_{2\dims}$.
}

\Remark{
  \Env{renumerate}{
    \item In the noiseless setting with $\epsilon = 0$, the above result
      immediately implies perfect recovery of all group-sparse signals as the
      $\dims$-term approximation error $\ktermerror{\dims}(\xorg)_{\partI,1}$
      vanishes as soon as $\xorg \in \sigset$.
    \item If desirable, it is also possible to characterize the recovery
      quality in terms of the group $\ell_{\partI,1}$-norm in which case one
      obtains
      \Equation*{
        \bnorm{\xorg - \opt{\vmx}}{1} \leq \tilde{C}
        \serror{\dims}(\xorg)_{\partI,1} + \tilde{D}\sqrt{\dims}\epsilon
      }
      for $\tilde{C},\tilde{D} > 0$ which still only depend on
      $\delta_{2\dims}$. {This is a conclusion of \blaref{thm:group_rip_guarantee}. See  \blaref{app:proof_thm_group_rip} for more discussions.
      }
  }
}

\subsection{Main Results}

Before stating our main result, we first recall the definition of subgaussian
random variables.

\Definition[Subgaussian random variable]{
  A zero mean random variable $X$ is called \emph{subgaussian} if there exists
  a constant $C>0$ such that
  \Equation*{
    \E(\exp(X^2/C^2))\leq 2.
  }
  The subgaussian norm of $X$, also known as Orlicz norm of $X$, is defined by
  \Equation*{
    \subgnorm{X}=\inf\set{C>0}[\E(\exp(X^2/C^2))\leq 2].
  }
}

At this point we are ready to state the main result of this paper.

\Theorem{\label{thm:block_diagonal_group_rip}
  Let $\vmA = \diag{\vmPhi_l}_{l=1}^\numO \in \R^{\dimy\numO \times \dimx\numO}$
  be a block diagonal random matrix with subgaussian blocks $\vmPhi_l$ whose
  entries are independent subgaussian zero-mean unit-variance random variables
  with subgaussian norm $\tau$. Let further $\vmPsi \in \Ugroup(\dimx\numO)$ be
  a unitary matrix. Then with probability at least $1 - \eta$, the matrix
  $\dimy^{-1/2}\vmA\vmPsi$ satisfies the \acl{GRIP} of order $\dims$ \wrt the
  group partition $\partI$, and $\delta_\dims \leq \delta$ if
  \Equation*{
    \dimy &\gtrsim_\tau \delta^{-2} \Big[\dims \coherence^2
    \log(\dimxtot)\log(\dims)^2 \parens{\log(\numG) +
    \dimg\log(\dims/\coherence)} + \log(\eta^{-1})\Big],
  }
  where
  \Equation*{
    \coherence = \coherence(\vmPsi)
    := \min\set{\sqrt{\dimx} \max_{i \in [\dimxtot]}{\bnorm{\vmpsi_i}{\infty}},
    1},
  }
  and $\vmpsi_i \in \C^\dimxtot$ denotes the $i$-th row of $\vmPsi$.
}

In the second acquisition model in which we assume that every sensor is
equipped with a copy of the same (random) measurement matrix $\vmPhi_l = \vmPhi$ for all $l \in [\numO]$, the coherence parameter $\coherence(\vmPsi)$ introduced
above is replaced by another parameter of the sparsity basis. To that end, we
introduce the following notation. Given a sparsity basis matrix $\vmPsi \in
\Ugroup(\dimxtot)$, denote by $\vmPsi_l \in \C^{\dimx \times \dimx\numO}$ the
$l$-th partial basis expansion matrix such that $\vmPsi =
\transp{(\transp{\vmPsi_1}, \ldots, \transp{\vmPsi_\numO})}$. With this
definition, the following result establishes the \ac{GRIP} for block diagonal
subgaussian random matrices with constant block-diagonal.

\Theorem{\label{thm:same_block_diagonal_group_rip}
  Under the conditions of \blaref{thm:block_diagonal_group_rip}, assume that
  ${\vmPhi}_{l}=\vmPhi$ for all $l\in[\numO]$ where the entries of $\vmPhi$
  are independent subgaussian zero-mean unit-variance random variables
  with subgaussian norm $\tau$. Then with probability at least $1 - \eta$, the
  matrix $\dimy^{-1/2}\vmA\vmPsi$ satisfies the \acl{GRIP} of order $\dims$
  \wrt the group partition $\partI$, and $\delta_\dims \leq \delta$ if
  \Equation*{
    \dimy &\gtrsim_\tau \delta^{-2} \Big[\dims \orthoconst^2
    \log(\dimxtot)\log(\dims)^2 \parens{\log(\numG) +
    \dimg\log(\dims/\orthoconst)}  + \log(\eta^{-1})\Big],
  }
  where
  \Equation*{
    &\orthoconst = \orthoconst(\vmPsi) := \min\set{
    \sqrt{\dimg} \max_{i \in [\dimxtot]}{\opnorm{\Vtilde(\vme^i)}},
    \sqrt{\numO} \max_{\substack{l \in [\numO], \\ i \in
    [\numG]}}{\opnorm{(\vmPsi_l)_{\partI_i}}}},
  }
  with $\vme^i$ denoting the $i$-th canonical unit vector and
  \Equation*{
    \Vtilde(\vmx) := \PMatrix{
      \transp{(\vmPsi_1\vmx)} \\
      \vdots \\
      \transp{(\vmPsi_\numO\vmx)}
    } \in \C^{\numO \times \dimx}.
  }
}
In the next sections, we briefly comment on a few observations of our attained
bound.

\subsection{{Discussions}}
\label{sec:discussions}

{Since the parameter $\coherence(\vmPsi)$ plays a central role,
some comments are in order.
First, let us point out that with the trivial group partition $\partilde =
\set{\set{1}, \ldots, \set{\dimxtot}}$, the parameter
$\coherence[\partilde](\vmPsi)$ reduces to the coherence parameter (up to
scaling by $\sqrt{\numO}$) considered in \cite{Eftekhari2012TheRI}.
In that case, the term $\Max[i \in [\dimxtot]]{\norm{\vmpsi_i}_\infty}$
corresponds to the constant associated with the \acl{BOS} generated by the
columns of the unitary matrix $\vmPsi$ as defined in
\cite[\chapName~12]{Foucart2013mathematicalIC}.
In general, the term $\coherence(\vmPsi)$ measures how coherent the sparsity
basis is with the canonical basis for $\C^\dimxtot$.
For instance, we clearly have $\coherence(\Id_\dimxtot) = 1$.
At the other end of the spectrum, we have for the orthogonal \ac{DFT} matrix \blaref*{eqn:dft_matrix}
that $\coherence(\vmF_\dimxtot) = \min\smallset{\sqrt{\dimg/\numO}, 1}$ since
every entry of $\vmF_\dimxtot$ has constant modulus and hence
$\bnorm{\vmpsi_i}{\infty} = \sqrt{\dimg/\dimxtot} $ for all $ i \in [\dimxtot]$.
This implies that the bound on $\radOp(\opimg)$ becomes more effective the
more sensors one considers.}

{
Consider the general block diagonal setup first in which every sensor is
equipped with an independent copy of a subgaussian random matrix.
For simplicity, we choose the failure
probability $\eta$ in \blaref{thm:block_diagonal_group_rip} such that the
condition on $\dimy$ simplifies to
\Equation{\label{eqn:simplified_measurement_bound_step_one}
  \dimy
  \gtrsim_\tau \delta^{-2} \dims \coherence(\vmPsi)^2
  \log(\dimxtot)\log(\dims)^2 \brackets{\log(\numG) + \dimg\log(\dims /
  \coherence(\vmPsi))}.
}
Moreover, we assume that the number of sensors $\numO$ exceeds the group size
$\dimg$.
As discussed above, the parameter
$\coherence(\vmPsi)$ ranges between the extreme points $\sqrt{\dimg/\numO}$
and $1$ corresponding to the choices $\vmPsi = \vmF_\dimxtot$ and $\vmPsi =
\Id_\dimxtot$, respectively.
We may therefore also lower bound $\coherence(\vmPsi)$ by $\sqrt{\dimg/\numO}$
in the last $\log$-factor of
\blaref*{eqn:simplified_measurement_bound_step_one}, which yields
\Equation{\label{eqn:simplified_measurement_bound_II}
  \dimy
  \gtrsim_\tau \delta^{-2} \dims \coherence(\vmPsi)^2
  \log(\dimxtot)\log(\dims)^2 \brackets{\log(\numG) + \dimg\log\parens{
    \frac{\dims\numO}{\dimg}}}.
}
For $\vmPsi = \vmF_\dimxtot$, this shows that the number of measurements per
sensor decreases almost linearly in $\numO$.
This in turn implies that roughly the same recovery fidelity can be maintained
if the number of measurements per sensor is reduced by adding more sensors to
the acquisition system.
However, since each sensor takes fewer samples in this scenario, this
ultimately results in a net gain since the energy consumption per sensor is
reduced.
On the other hand, if target signals are group-sparse \wrt the canonical basis,
such a reduction does not seem possible.
This is due to the fact that in the worst case scenario, all active groups
might be restricted to a single chunk $\vmx_l$.
In this case, each measurement operator $\vmPhi_l$ has to act as a \ac{GRIP}
matrix.
This drawback is fundamental to the acquisition model and cannot be overcome
by a refined proof technique. 
}
{ A similar discussion applies to the coherence term $\orthoconst$. As we have seen above, the term 
$\Max[i \in [\dimxtot]]{\opnorm{\Vtilde(\vme^i)}}$ is between $1/\sqrt{L}$ and $1$. For $\vmPsi$ as the identity matrix, $\Max[i \in [\dimxtot]]{\opnorm{\Vtilde(\vme^i)}}$ is equal to $1$. The coherence value $\orthoconst$ in this case is given by  to $\min\{\sqrt{g},\sqrt{L}\}$. When $g$ is larger than the number of sensors, then the overall scaling is not desirable. Note that our problem is reduced to the result in \cite{Eftekhari2012TheRI}  by choosing $g=1$.  On the other hand, consider an orthonormal basis that is drawn uniformly from the orthogonal group. It has been shown  in \cite[Lemma 1]{Eftekhari2012TheRI} that the term 
$\Max[i \in [\dimxtot]]{\opnorm{\Vtilde(\vme^i)}}$ scales as $1/\sqrt{L}$. In this case, the coherence term $\orthoconst$ is upper bounded by $\sqrt{g/L}$ and can be decreased by increasing $L$. This is similar to what we observed for $\mi_I(\vmPsi)$. Namely, the number of measurements per sensor decreases almost linearly in $\numO$.
}

\subsection{Connection to Sparse Vector Recovery}

First, let us observe what happens when the maximum group size $\dimg$ tends
to 1, and therefore $\numG = \dimxtot$. For simplicity, we choose the failure
probability $\eta$ in \blaref{thm:block_diagonal_group_rip} such that the
condition on $\dimy$ simplifies to
\Equation{\label{eqn:simplified_measurement_bound}
  \dimy \gtrsim c_\tau \delta^{-2} \dims \coherence(\vmPsi)^2
  \log(\dimxtot)\log(\dims)^2 \parens{\log(\numG) + \dimg\log(\dims)}
}
where we also dropped the coherence parameter $\coherence(\vmPsi) \leq 1$ in
the last $\log$-factor. When the group size tends to $1$, and we are dealing
with sparse rather than group-sparse vectors as considered in
\cite{Eftekhari2012TheRI} and \cite{Chun2016UniformRF}, the required number of
measurements for $\numG = \dimxtot$ reduces to
\Equation*{
  \dimy \gtrsim c_\tau \delta^{-2} \dims
  \coherence(\vmPsi)^2\log(\dimxtot)\log(\dims)^2 \parens{\log(\dimxtot) +
  \log(\dims)}.
}
Since $\dims \leq \dimxtot$, it consequently suffices to choose
\Equation*{
  \dimy \gtrsim c_\tau \delta^{-2} \dims
  \coherence(\vmPsi)^2\log(\dimxtot)^2\log(\dims)^2.
}
Recalling the definition of the coherence parameter
\Equation*{
  \coherence(\vmPsi) = \min\set{\sqrt{\dimx} \max_{i \in
  [\dimxtot]}{\bnorm{\vmpsi_i}{\infty}}, 1},
}
we have for $\partI = \set{\set{1}, \ldots, \set{\dimxtot}}$ that
$\bnorm{\cdot}{\infty} = \norm{\cdot}_\infty$, and therefore
\Equation*{
  \coherence(\vmPsi) = \frac{1}{\sqrt{\numO}} \min\set{\sqrt{\dimxtot} \Max[i
  \in [\dimxtot]]{\norm{\vmpsi_i}_\infty}, \sqrt{\numO}} =:
  \frac{1}{\sqrt{\numO}} \mu(\vmPsi)
}
where $\mu(\vmPsi)$ denotes a rescaled coherence parameter in accordance with
the definition used by Eftekhari \etal (\cf \eqnName~$(5)$ in
\cite{Eftekhari2012TheRI}). This now implies
\Equation*{
  \dimy\numO \gtrsim_\tau \delta^{-2} \dims \mu(\vmPsi)^2
  \log(\dimxtot)^2\log(\dims)^2
}
which is precisely the statement of \thmName~1 in \cite{Eftekhari2012TheRI}.
The same argument yields the specialization to the situation in which each
sensor is equipped with the same random matrix $\vmPhi \in \R^{\dimy \times
\dimx}$. As we will discuss in \blaref{sec:shared_sensing_matrix}, the
parameter $\orthoconst(\vmPsi)$ \wrt the trivial group partition $\partI =
\set{\set{1}, \ldots, \set{\dimxtot}}$ reduces to
\Equation*{
  \orthoconst(\vmPsi) = \Max[i \in [\dimxtot]]{\opnorm{\Vtilde(\vme^i)}}.
}
Defining the so-called \textit{block-coherence} parameter $\gamma(\vmPsi) :=
\sqrt{\numO} \orthoconst(\vmPsi)$ to borrow terminology from Eftekhari \etal
(\cf \cite[\eqnName~$(9)$]{Eftekhari2012TheRI}), this yields the condition
\Equation*{
  \dimy\numO \gtrsim_\tau \delta^{-2} \dims \gamma(\vmPsi)^2
  \log(\dimxtot)^2\log(\dims)^2
}
which reproduces the statement of \thmName~2 in \cite{Eftekhari2012TheRI}.

\subsection{Comparison to Dense Measurement Matrices}

As alluded to in the introduction, it is by now a well-established fact that
{$\Omega({\dims\log(\dimx/\dims)})$
}
nonadaptive measurements based on subgaussian
random ensembles are sufficient to stably reconstruct sparse or compressible
vectors from their linear projections. Moreover, this bound is fundamental in
that it is known to be optimal among all encoder-decoder pairs $(\vmA,\Delta)$
with $\vmA \in \C^{\dimy \times \dimx}$ and decoding maps
$\function{\Delta}{\C^\dimy}{\C^\dimx}$ such that
\Equation*{
  \norm{\vmx - \Delta(\vmA\vmx)}_2 \leq \frac{C}{\sqrt{\dims}}
  \serror{\dims}(\vmx)_1 \quad\Forall \vmx \in \C^\dimx
}
for $C > 0$ \cite[\chapName~10]{Foucart2013mathematicalIC}. Such a fundamental
lower bound on the required number of measurements was recently also
established for the case of block-sparse vectors by Dirksen and Ullrich
\cite{Dirksen2017GelfandNR} (see also \cite[\thmName~2.4]{Ayaz2014UniformRO}).
In particular, using new results on Gelfand numbers, the authors show that
stability results of the form
\Equation*{
  \norm{\vmx - \Delta(\vmA\vmx)}_2 \leq \frac{C}{\sqrt{\dims}}
  \serror{\dims}(\vmx)_{\partI,1} \quad \Forall \vmx \in \C^\dimxtot
}
for arbitrary encoder-decoder pairs $(\vmA, \Delta)$ require at least
\Equation*{
  \dimytot \geq c_1(\dims\log(e\numG/\dims) + \dims\dimg) \quad\text{with}\quad
  \dims > c_2
}
measurements where the constants $c_1$ and $c_2$ only depend on $C > 0$ (\cf
\cite[\corName~1.2]{Dirksen2017GelfandNR}). Perhaps most surprisingly about
this result is the linear dependence on the total number of nonzero
coefficients $\dims\dimg$.
In light of
\blaref{eqn:simplified_measurement_bound}, we also recover this scaling
behavior in the total number of measurements $\dimytot$ for the block diagonal
measurement setup, albeit with the additional logarithmic factor in $\dims$
which we conjecture to be an artifact of the proof technique employed in
\blaref{sec:metric_entropy_bound}. The other polylogarithmic factors, as well
as the dependence on $\coherence(\vmPsi)$, on the other hand, are due to
the particulars of the measurement setup compared to the situation in which we
employ one densely populated measurement matrix to observe the entire signal.
Whether these factors can be improved any further remains an open problem.

\subsection{Connection to Distributed Sensing}
\label{sec:connection_to_dcs}

As mentioned in the introduction, the measurement model
\blaref*{eqn:measurements} frequently appears in the context of recovering
multiple versions of a vector sharing a common low-complexity structure. This
model appears for instance in the context of distributed sensing where one aims
to estimate the structure of a ground truth signal observed by spatially
distributed sensors which each observe a slightly different version of the
signal due to channel propagation effects.

Another classic example is that of the so-called \acf{MMV} model in which a
single sensor acquires various temporal snapshots of a signal whose
low-complexity structure is assumed to be stationary\footnote{In particular,
this model assumes the sparse support set to be constant, while amplitudes and
phases of the coefficients of each vector are allowed to change between
different observations.} with the intent of reducing the influence of
measurement noise in a single-snapshot model. This particular model can be cast
in the setting of \blaref{sec:shared_sensing_matrix} where we interpret each
observation in the \ac{MMV} model as an independent observation by a distinct
sensor equipped with the same measurement matrix $\vmPhi \in \R^{\dimy \times
\dimx}$.

Assuming that the ground truth signal is $\dims$-sparse, we can interpret both
situations as trying to recover an $\dims$-group-sparse vector \wrt the group
partition $\partI = \set{\partI_1, \ldots, \partI_\dimx}$ with
\Equation{\label{eqn:dist_group_partition}
  \partI_i = \set{i, \dimx+i, \ldots, (\numO-1)\dimx + i}.
}
In both situations, we assume that each signal $\vmz^l =
\widetilde{\vmPsi}\vmx^l \in \C^\dimx$ is sparse in the same basis
$\widetilde{\vmPsi} \in \Ugroup(\dimx)$. We can therefore choose $\vmPsi =
\diag*{\widetilde{\vmPsi}}_{l=1}^\numO \in \Ugroup(\dimxtot)$ in
\blaref{thm:block_diagonal_group_rip}. This setup, however, is not able to cope
with certain adversarial vectors. More precisely, due to the particular group
partition structure, the knowledge about the periodicity in the support
structure can not necessarily be exploited in all recovery scenarios. To see
this, consider the situation in which only a single vector $\vmx^l$ is
different from $\vmzero$. The vector $\vmx = \transp{(\transp{\vmzero}, \ldots,
\transp{\vmzero}, \transp{(\vmx^l)}, \transp{\vmzero}, \ldots,
\transp{\vmzero})}$ is then by definition $\dims$-group-sparse (\wrt the group
partition $\partI$) if $\vmx^l$ is $\dims$-sparse. Regardless of the sparsity
basis $\widetilde{\vmPsi} \in \Ugroup(\dimx)$, only the vector $\vmy^l$ carries
information about $\vmx^l$ which implies that each matrix $\vmPhi_l$ should
satisfy the classical \acl{RIP} to recover $\vmx$. This happens with high
probability as soon as $\dimy = {\Omega}({\dims \log(\dimx/\dims)})$. In this case,
instead of solving \blaref{prob:group_qcbp} directly, it is more favorable to
solve for each $l \in [\numO]$ the problem
\Equation{\label{prob:degenerate_qcbp}
  \tagprob
  \MinProblem{\norm*{\vmx^l}_1}{\norm*{\vmy^l -
  \vmPhi_l\widetilde{\vmPsi}\vmx^l}_2 \leq \epsilon.}
}
Unfortunately, this behavior is not accurately captured by
\blaref{thm:block_diagonal_group_rip} since we have by
\blaref{eqn:simplified_measurement_bound} with $\numG = \dimx$ and $\dimg =
\numO$ that
\Equation*{
  \dimy \gtrsim c_\tau \delta^{-2} \coherence(\vmPsi)^2 \dims\log(\dimxtot)
  \log(\dims)^2(\log(\dimx) + \numO\log(\dims)).
}
This predicts a much worse scaling behavior than what is required to solve
$\numO$ separate problems of the form \blaref*{prob:degenerate_qcbp}. The
problem is ultimately rooted in the fact that independent of
$\widetilde{\vmPsi} \in \Ugroup(\dimx)$, only the measurements $\vmy^l$ carry
information about $\vmx^l$.

Note that such adversarial situations had previously been discussed by van den
Berg and Friedlander \cite{Berg2009JointsparseRF} who consider sufficiency
conditions for noiseless joint-sparse recovery based on dual certificates.
Instead of considering signals with only one $\dims$-sparse nonzero signal
$\vmx^l$, they consider signals $\vmx$ in which every $\vmx^l$ is at
most $1$-sparse with $\supp(\vmx^l) \neq \supp(\vmx^{l'})$ for any $l \neq l'$.
In this setting, they show that there are signals $\xorg \in \R^\dimxtot$
which---given the linear measurements $\vmy = \diag*{\vmPhi}_{l=1}^\numO
\xorg$---can provably be recovered by the program
\Equation*{
  \MinProblem{\norm{\vmx}_1}{\vmy = \diag{\vmPhi}_{l=1}^\numO \vmx}
}
but not via group $\ell_{\partI,1}$-minimization, \ie, as solutions of
\blaref{prob:group_qcbp} with $\vmPsi = \Id_\dimxtot$, and $\epsilon = 0$.

The problem of \acl{DCS} was also recently addressed in the context of
quantized \acl{CS} with binary observations by Maly and Palzer
\cite{Maly2018AnalysisOH} who impose an additional norm constraint on each
signal to avoid that $\vmx^l = \vmzero$. However, even with this modified
signal model, the adversarial example discussed above still applies if one
signal $\vmx^l$ is exactly $\dims$-sparse, while any other signal $\vmx^{l'}$
with $l' \neq l$ is $1$-sparse with the entire signal energy concentrated on
the same coordinate in each vector $\vmx^{l'}$. The resulting signal is
therefore $\dims$-group-sparse as in the previous example. In that case, each
measurement vector $\vmy^{l'}$ only carries information about a single nonzero
coordinate of $\vmx^l$ which implies that each $\vmPhi_l$ must itself be able
to recover every $(\dims-1)$-sparse vector for the entire vector $\vmx$ to be
recovered as desired.

To summarize, without further restrictions on the particular signal model, it
is not clear how adversarial examples as discussed above can be dealt with in
order to obtain nontrivial uniform recovery guarantees. However, the conclusion
of the work in \cite{Eftekhari2012TheRI} and our results is that sparsity or
group-sparsity in a nonlocalized unitary basis such as the \ac{DFT} basis bears
the potential to reduce the number of measurements required for stable and
robust signal recovery by distributing the energy of nonzero coefficients
across the entire signal support.
As pointed out above, however, this requires that the sparsity basis of $\vmx$
does \emph{not} take the form of a block diagonal unitary matrix.

\section{The \texorpdfstring{\Acs{GRIP}}{Group-RIP} for Block Diagonal
Matrices}
\label{sec:proof_bdgrip}

In this section, we establish the \ac{GRIP} for general subgaussian block
diagonal matrices.

\subsection{Restricted Isometries and Suprema of Chaos Processes}

We will make use of the following powerful bound on the suprema of chaos
processes first established in \cite[\thmName~3.1]{Krahmer2012SupremaOC} to
demonstrate that the block diagonal matrix $\vmA\vmPsi \in \C^{\dimytot \times
\dimxtot}$ satisfies the \acl{GRIP} with high probability on the draw of
$\vmA$. The same technique was also employed in \cite{Eftekhari2012TheRI} to
prove the canonical \acl{RIP} for block diagonal matrices consisting of
subgaussian blocks. In the present work, we make use of an improved version of
the bound due to Dirksen \cite{Dirksen2013TailBV}. Before stating the result,
we first define the following objects. Let $\opimg \subset \C^{m \times n}$ be
a bounded set. Then the
radii of $\opimg$ \wrt the Frobenius and operator norm are defined as
\Equation*{
  \radFrob(\opimg) = \sup_{\vmGamma \in \opimg}{\Fnorm{\vmGamma}}
  \quad\text{and}\quad
  \radOp(\opimg) = \sup_{\vmGamma \in \opimg}{\opnorm{\vmGamma}},
}
respectively. Lastly, we require the so-called $\gamFunc$-functional of
$\opimg$ \wrt the operator norm.

\Definition{
  An admissible sequence of a metric space $(T,d)$ is a collection
  $\set{T_r\subset T: r\geq 0}$ where $\card{T_r}\leq 2^{2^r}$ for every $r\geq
  1$ and $\card{T_0}=1$. The $\gamFunc$ functional is defined by
  \Equation*{
    \gamFunc(T,d)=\inf \sup_{t\in T}\sum_{r=0}^\infty 2^{r/2}d(t,T_r),
  }
  where the infimum is taken over all admissible sequences.
}

It is generally difficult to characterize $\gamFunc$ directly. To estimate
$\gamFunc$, it is therefore customary to appeal to a classical result due to
Talagrand which bounds $\gamFunc(\opimg, \opnorm{\cdot})$ in terms of the
following entropy integral of the metric space\footnote{The metric on $\opimg$
is the one induced by the norm $\opnorm{\cdot}$.} $(\opimg,\opnorm{\cdot})$
\cite{Talagrand2010TheGC}:
\Equation{\label{eqn:entropy_integral}
  \gamFunc(\opimg, \opnorm{\cdot}) \lesssim \int_0^\infty \sqrt{\log
  \covnum(\opimg, \opnorm{\cdot}, \varepsilon)} \de{\varepsilon}
}
where $\covnum$ denotes the internal covering number, \ie, the cardinality of
the smallest subset $\net \subset \opimg$ such that every point in $\opimg$ is
at most $\varepsilon$ apart from $\net$ \wrt the operator norm
$\opnorm{\cdot}$. Mathematically, $\net \subset \opimg$ is called an
$\varepsilon$-net of $\opimg$ if for all $\vmGamma \in \opimg$, there exists $\vmGamma_0 \in \net : \opnorm{\vmGamma - \vmGamma_0} \leq \varepsilon$ with
$\covnum(\opimg,\opnorm{\cdot},\varepsilon) = \card{\net}$ if $\net$ is the
smallest such net. Note that the integrand of the entropy integral
\blaref*{eqn:entropy_integral} vanishes as soon as $\varepsilon \geq
\radOp(\opimg)$ since $\opimg$ can then be covered by a single ball
$\ball{2\to2}^{m \times n}$ centered at an (arbitrary) element of $\opimg$.

\Theorem[\cite[\thmName~6.5]{Dirksen2013TailBV}]{\label{thm:chaos_bound}
  Let $\opimg$ be a matrix set, and denote by $\vmxi$ an isotropic
  unit-variance subgaussian\footnote{The subgaussian property readily implies
  that $\vmxi$ is centered.} random vector with subgaussian norm $\tau$. Then,
  for $u \geq 1$,
  \Equation*{
    \P\parens{\sup_{\vmGamma \in \opimg}{\abs{\norm{\vmGamma\vmxi}_2^2 -
    \E\norm{\vmGamma\vmxi}_2^2}} \geq c_\tau E_u} \leq e^{-u}
  }
  where
  \Equation*{
    E_u &= \gamFunc(\opimg, \opnorm{\cdot})^2 +
    \radFrob(\opimg)\gamFunc(\opimg,\opnorm{\cdot}) + \sqrt{u}\radFrob(\opimg)\radOp(\opimg) + u \radOp(\opimg)^2,
  }
  and $c_\tau$ is a constant that only depends on $\tau$.
}

\subsection{Chaos Process for Block-Diagonal
\texorpdfstring{\Acs{GRIP}}{Group-RIP} Matrices}

In order to apply \blaref{thm:chaos_bound} to estimate the probability that
$\vmA\vmPsi$ as defined in \blaref{eqn:measurements} satisfies the \acl{GRIP},
first note that we can equivalently express the \acs{GRIP} condition in
\blaref{eqn:group_rip}
as
\Equation*{
  \abs{\frac{\norm{\vmA\vmPsi\vmx}_2^2}{\norm{\vmx}_2^2} - 1} \leq \delta \quad {\forall\vmx \in \sigset \setminus \set{\vmzero}}.
}
With the definition of the set
\Equation*{
  \sigsetO := \sigset \cap \sphere{\dimxtot-1} = \set{\vmx \in
  \sphere{\dimxtot-1}}[\bnorm{\vmx}{0} \leq \dims]
}
of $\dims$-group-sparse vectors on the unit Euclidean sphere, we may therefore
write the \acl{GRIC} of $\vmA$ as
\Equation{\label{eqn:group_ric}
  \delta_\dims = \sup_{\vmx \in \sigsetO}{\abs{\norm{\vmA\vmPsi\vmx}_2^2 - 1}}.
}
Next, we transform the above expression into the form required by
\blaref{thm:chaos_bound}, \ie, we rewrite the equation so that the supremum is
taken over a matrix set. To that end, recall the definition of the partial
basis expansion matrices $\vmPsi_l \in \C^{\dimx \times \dimx\numO}$ with
$\vmPsi = \transp{(\transp{\vmPsi_1}, \ldots, \transp{\vmPsi_\numO})}$. In
light of \blaref{eqn:measurements}, we may now express the $l$-th measurement
vector $\vmy^l \in \C^\dimy$ of $\vmy \in \C^{\dimy\numO}$ as
\Equation*{
  \vmy^l &= \vmPhi_l\vmPsi_l\vmx = \PMatrix{
    \inner{(\vmPhi_l)_1}{\vmPsi_l\vmx} \\
    \vdots \\
    \inner{(\vmPhi_l)_\dimy}{\vmPsi_l\vmx}
  } \\
  &= \underbrace{\PMatrix{
    \transp{(\vmPsi_l\vmx)} & & \\
    & \ddots & \\
    & & \transp{(\vmPsi_l\vmx)}
  }}_{=: V_l(\vmx) \in \C^{\dimy \times \dimy\dimx}} \cdot
  \underbrace{\PMatrix{(\vmPhi_l)_1 \\ \vdots \\ (\vmPhi_l)_\dimy}}_{=: \vmxi^l
  \in \R^{\dimy\dimx}}
}
where $(\vmPhi_l)_i \in \C^{\dimx}$ denotes the $i$-th row of the matrix
$\vmPhi_l$. If the blocks $\vmPhi_l$ are populated by independent copies of a
$\tau$-subgaussian random variable with unit-variance, then
the vector $\vmxi = \transp{(\transp{(\vmxi^1)}, \ldots,
\transp{(\vmxi^\numO)})}$ is a unit-variance $\tau$-subgaussian random vector.
Defining the linear operator $\function{V}{\C^{\dimx\numO}}{\C^{\dimy\numO \times
\dimy\dimx\numO}}$ with
\Equation{\label{eqn:V_operator}
  \vmx \mapsto V(\vmx) = \diag{V_l(\vmx)}_{l=1}^\numO,
}
we therefore have $\vmA\vmPsi\vmx \eqdist V(\vmx) \vmxi$ where $\eqdist$
denotes equality in distribution. Now note that
\Equation*{
  \E\norm{\vmA\vmPsi\vmx}_2^2 =
  \adj{\vmx}\adj{\vmPsi}\E\brackets{\transp{\vmA}\vmA}\vmPsi\vmx = \dimy
  \norm{\vmx}_2^2
}
which follows from the fact that the rows of the matrices $\vmA_l$ are
independent unit-variance random $\dimy$-vectors with independent entries, as
well as from unitarity of $\vmPsi$. With \blaref{eqn:group_ric}, the \acl{GRIP}
of the matrix $1/\sqrt{\dimy}\vmA\vmPsi$ can therefore be expressed as
\Equation*{
  \delta_\dims\parens{\frac{1}{\sqrt{\dimy}}\vmA\vmPsi} &=
  \sup_{\vmx \in \sigsetO}{\abs{\norm{\frac{1}{\sqrt{\dimy}}\vmA\vmPsi\vmx}_2^2
  - 1}} \\
  &= \sup_{\vmx \in
  \sigsetO}{\abs{\frac{1}{\dimy}\norm{\vmA\vmPsi\vmx}_2^2 -
  \frac{1}{\dimy}\dimy\norm{\vmx}_2^2}} \\
  &= \frac{1}{\dimy} \sup_{\vmx \in \sigsetO}{\abs{\norm{\vmA\vmPsi\vmx}_2^2 -
  \E\norm{\vmA\vmPsi\vmx}_2^2}} \\
  &\eqdist \frac{1}{\dimy} \sup_{\vmx \in
  \sigsetO}{\abs{\norm{V(\vmx)\vmxi}_2^2 - \E\norm{V(\vmx)\vmxi}_2^2}} \\
  &= \frac{1}{\dimy} \sup_{\vmGamma \in \opimg}{\abs{\norm{\vmGamma\vmxi}_2^2 -
  \E\norm{\vmGamma\vmxi}_2^2}}
}
where we set $\opimg := V(\sigsetO) = \set{V(\vmx)}[\vmx \in \sigsetO]$. In
order to apply \blaref{thm:chaos_bound}, it remains to estimate the radii of
$\opimg$ \wrt the Frobenius and operator norm, respectively, as well as to
compute the $\gamFunc$-functional of $\opimg$ \wrt $\opnorm{\cdot}$. These
issues are addressed in the next two sections.

\subsection{Radii Estimates}
We begin with the estimation of $\radFrob(\opimg)$. To that end, first note
that
\Equation*{
  \Fnorm{V(\vmx)}^2 &= \Fnorm{\diag{V_l(\vmx)}_{l=1}^\numO}^2 =
  \sum_{l=1}^\numO \Fnorm{V_l(\vmx)}^2 \\
  &= \sum_{l=1}^\numO \dimy \norm{\vmPsi_l\vmx}_2^2 = \dimy
  \norm{\vmPsi\vmx}_2^2 = \dimy \norm{\vmx}_2^2.
}
Since $\sigsetO \subset \sphere{\dimxtot-1}$, this immediately implies
\Equation*{
  \radFrob(\opimg) = \sup_{\vmGamma \in \opimg}{\Fnorm{\vmGamma}} = \sup_{\vmx
  \in \sigsetO}{\Fnorm{V(\vmx)}} = \sqrt{\dimy} \sup_{\vmx \in
  \sigsetO}{\norm{\vmx}_2} = \sqrt{\dimy}.
}

In order to estimate the radius $\radOp(\opimg)$, we require a simple
generalization of Hölder's inequality to group $\ell_{\partI,p}$-norms on
$\C^\dimxtot$ as defined in \blaref{def:group_lp_norms}. We state here a
specialization to the conjugate pair $p=1$, $q=\infty$.
\Lemma{\label{lem:group_hoelder}
  Let $\vma,\vmb \in \C^\dimxtot$, and let $\partI$ be a group partition of
  $[\dimxtot]$. Then
  \Equation*{
    \abs{\inner{\vma}{\vmb}} \leq \bnorm{\vma}{1} \cdot \bnorm{\vmb}{\infty}
  }
  where $\inner{\cdot}{\cdot}$ denotes the bilinear form $\inner{\vma}{\vmb} =
  \sum_{i=1}^\dimxtot a_i b_i$ on $\C^\dimxtot$.
}
\Proof{
  By the triangle and Hölder's inequality, we have
  \Equation*{
    &\Phantom{=}\abs{\inner{\vma}{\vmb}} \\
    &= \abs{\sum_{i=1}^\numG
    \inner{\vma_{\partI_i}}{\vmb_{\partI_i}}} \leq \sum_{i=1}^\numG
    \abs{\inner{\vma_{\partI_i}}{\vmb_{\partI_i}}} \leq \sum_{i=1}^\numG
    \norm{\vma_{\partI_i}}_2 \cdot \norm{\vmb_{\partI_i}}_2 \\
    &\leq \sum_{i=1}^\numG \norm{\vma_{\partI_i}}_2 \cdot \max_{j \in
    [\numG]}{\norm{\vmb_{\partI_j}}_2} = \bnorm{\vma}{1} \cdot
    \bnorm{\vmb}{\infty}.
  }
}

We proceed as before and compute
\Equation{
  \opnorm{V(\vmx)} &= \opnorm{\diag{V_l(\vmx)}_{l=1}^\numO} = \max_{l \in
  [\numO]}{\opnorm{V_l(\vmx)}} \notag \\
  &= \max_{l \in [\numO]}{\opnorm{V_l(\vmx)\adj{V_l(\vmx)}}^{1/2}} = \max_{l
  \in [\numO]}{\norm{\vmPsi_l\vmx}_2} \label{eqn:induced_norm}
}
where the second step follows from the fact that the operator norm of a block
diagonal matrix corresponds to the maximum operator norm of the individual
blocks. The last step follows because $V_l(\vmx)\adj{V_l(\vmx)}$ is a diagonal
matrix with $\dimy$ copies of $\norm{\vmPsi_l\vmx}_2^2$ on its diagonal whose
largest singular value is simply $\norm{\vmPsi_l\vmx}_2^2$. Next, we invoke the
bound $\norm{\vmx}_2 \leq \sqrt{n} \norm{\vmx}_\infty$ for $\vmx \in \C^n$,
followed by an application of \blaref{lem:group_hoelder}. This yields
\Equation*{
  \norm{\vmPsi_l\vmx}_2 &\leq \sqrt{\dimx} \norm{\vmPsi_l\vmx}_\infty =
  \sqrt{\dimx} \max_{i \in [\dimx]}{\abs{\inner{(\vmPsi_l)_i}{\vmx}}} \\
  &\leq \sqrt{\dimx} \max_{i \in [\dimx]}{\bnorm{(\vmPsi_l)_i}{\infty} \cdot
  \bnorm{\vmx}{1}}
}
where $(\vmPsi_l)_i$ denotes the $i$-th row of $\vmPsi_l$. Overall, we find
\Equation*{
  \opnorm{V(\vmx)} &\leq \sqrt{\dimx} \bnorm{\vmx}{1} \max_{\substack{l \in
  [\numO], \\ i \in [\dimx]}}{\bnorm{(\vmPsi_l)_i}{\infty}} \\
  &= \sqrt{\dimx} \bnorm{\vmx}{1} \max_{i \in
  [\dimxtot]}{\bnorm{\vmpsi_i}{\infty}}
}
where $\vmpsi_i \in \C^\dimxtot$ denotes the $i$-th row of $\vmPsi$. This bound
is less effective for instance when $\vmPsi = \Id_\dimxtot$ but more so when
$\vmPsi$ corresponds to a \ac{DFT} matrix, \ie,
\Equation{\label{eqn:dft_matrix}
  \vmPsi = \vmF_\dimxtot = \frac{1}{\sqrt{\dimxtot}} \parens{e^{i2\pi
  mn/\dimxtot}}_{0 \leq m,n \leq \dimxtot-1}.
}
For $\vmPsi = \Id_\dimxtot$, we have $\sqrt{\dimx} \max_{i \in
[\dimxtot]}{\bnorm{\vmpsi_i}{\infty}} = \sqrt{\dimx}$, whereas for $\vmPsi =
\vmF_\dimxtot$ we get $\sqrt{\dimx}\max_{i \in
[\dimxtot]}{\bnorm{\vmpsi_i}{\infty}} = \sqrt{\dimg/\numO}$ with $\dimg =
\max_{i \in [\numG]}{\card{\partI_i}}$ denoting the size of the largest
coefficient group. To obtain an effective bound in both situations, we
therefore also consider the simple bound
\Equation{
  \opnorm{V(\vmx)} &= \max_{l \in [\numO]}{\norm{\vmPsi_l\vmx}_2} \leq
  \norm{\vmPsi\vmx}_2 \notag \\
  &= \norm{\vmx}_2 = \bnorm{\vmx}{2} \leq \bnorm{\vmx}{1}
  \label{eqn:induced_norm_simple_bound}
}
which follows from $\norm{\cdot}_p \leq \norm{\cdot}_q$ for $p \geq q \geq 1$.
Combining both estimates, we arrive at
\Equation*{
  \radOp(\opimg) &= \sup_{\vmx \in \sigsetO}{\opnorm{V(\vmx)}} \\
  &\leq \sup_{\vmx
  \in \sigsetO}{\bnorm{\vmx}{1}} \min\set{\sqrt{\dimx} \max_{i \in
  [\dimxtot]}{\bnorm{\vmpsi_i}{\infty}}, 1} \\
  &\leq \sqrt{\dims} \min\set{\sqrt{\dimx} \max_{i \in
  [\dimxtot]}{\bnorm{\vmpsi_i}{\infty}}, 1} \\
  &=: \sqrt{\dims} \coherence(\vmPsi).
}

The last inequality holds since for $\vmx \in \sigsetO = \sigset \cap
\sphere{\dimxtot-1}$, we have
\Equation*{
  \bnorm{\vmx}{1} &= \sum_{i=1}^\numG \norm{\vmx_{\partI_i}}_2 \leq
  \parens{\sum_{i=1}^\numG \norm{\vmx_{\partI_i}}_2^2}^{1/2}
  \parens{\sum_{i=1}^\numG \binarize{\vmx_{\partI_i} \neq \vmzero}}^{1/2} \\
  &\leq \bnorm{\vmx}{2} \sqrt{\dims} = \sqrt{\dims}\norm{\vmx}_2 = \sqrt{\dims}
}
by the Cauchy-Schwarz inequality.

\subsection{Metric Entropy Bound}
\label{sec:metric_entropy_bound}

Establishing a bound on the $\gamFunc$-functional via
\blaref{eqn:entropy_integral} will proceed in two steps. At small scales, we
will estimate the covering number by means of a standard volume comparison
argument for norm balls covered in their respective metrics. At larger scales,
however, this bound will not be effective enough to yield optimal scaling
behavior in $\dims$. To circumvent the problem, we employ a variation of
Maurey's empirical method.

To start with, note that with $\norm{\vmx}_V := \opnorm{V(\vmx)}$, we have for
$u \geq 0$,
\Equation*{
  \covnum(\opimg, \opnorm{\cdot}, u) = \covnum(\sigsetO, \norm{\cdot}_V, u).
}
With this we decompose the metric entropy integral as
\Equation{
  &\Phantom{=}\int_0^{\radOp(\opimg)} \sqrt{\log\covnum(\opimg, \opnorm{\cdot},
  \varepsilon)} \de{\varepsilon} \notag \\
  &\leq \int_0^\lambda \sqrt{\log\covnum(\sigsetO, \norm{\cdot}_V,
  \varepsilon)} \de{\varepsilon}
  + \int_\lambda^{\sqrt{\dims} \coherence(\vmPsi)} \sqrt{\log\covnum(\sigsetO,
  \norm{\cdot}_V, \varepsilon)} \de{\varepsilon}
  \label{eqn:entropy_integral_parts}
}
where the parameter $\lambda \in [0, \sqrt{\dims} \coherence(\vmPsi)]$ will be
chosen later.
Next, we may express the set $\sigsetO = \sigset \cap \sphere{\dimxtot-1}$ of
$\dims$-group-sparse signals on the unit sphere as the union of
$\binom{\numG}{\dims}$ unit Euclidean spheres supported on $\dims$ groups of a
group partition $\partI$. Denote for $\Tset \subset \partI$ the coordinate
subspace of $\C^\dimxtot$ supported on the index set $\bigcup_{S \in \Tset} S
\subset [\dimxtot]$ by $\C_\Tset^\dimxtot$, \ie,
\Equation*{
  \C_\Tset^\dimxtot = \set{\vmx \in \C^\dimxtot}[\vmx_S = \vmzero
  \Forall S \notin \Tset].
}
Then we can write
\Equation*{
  \sigsetO = \bigcup_{\substack{\Tset \subset \partI, \\ \card{\Tset}=\dims}}
  (\sphere{\dimxtot-1} \cap \C_\Tset^\dimxtot)
  \subset \bigcup_{\substack{\Tset \subset \partI, \\ \card{\Tset}=\dims}}
  (\ball{2}^\dimxtot \cap \C_\Tset^\dimxtot).
}
The linear-algebraic dimension of the sets in this union is at most
$\dims\dimg$ where again $\dimg$ denotes the largest group of the partition
$\partI$ considered in $\Tset$. From the volume comparison argument for norm
balls covered in their associated metrics (see \eg \cite[Corollary
4.2.13]{Vershynin2018HighDP}), one has that $\covnum(\ball{\norm{\cdot}}^n,
\norm{\cdot}, t) \leq (1+2/t)^n$. With \blaref{eqn:induced_norm_simple_bound},
this yields for an arbitrary group index set $\Tset$ as above that
\Equation{
  \covnum(\sigsetO, \norm{\cdot}_V, u)
  &\leq \binom{\numG}{\dims} \covnum(\ball{2}^\dimxtot \cap \C_\Tset^\dimxtot,
  \norm{\cdot}_2, u/2) \notag \\
  &\leq \parens{\frac{e\numG}{\dims}}^\dims \parens{1 +
  \frac{4}{u}}^{2\dims\dimg} \label{eqn:covnum_bound_1}
}
where the factor $1/2$ in the covering radius of the first estimate is due to
the fact that the internal covering numbers are only almost increasing by
inclusion, \ie, if $U \subset W$, then $\covnum(U, \cdot, t) \leq \covnum(W,
\cdot, t/2)$ \cite[Exercise 4.2.10]{Vershynin2018HighDP}. The factor $2$ in the
exponent of the last estimate is due to the isomorphic identification of $\C^n$
with $\R^{2n}$. Finally, we invoked the standard bound $\binom{n}{k} \leq
(en/k)^k$ for binomial coefficients.

{To estimate $\covnum(\sigsetO, \norm{\cdot}_V,u)$ at higher scales, we develop
a variation on Maurey's empirical method, also known as \textit{Maurey's
lemma} \cite{carl1985inequalities}. 
In general, Maurey's lemma is concerned with the following question.
Given a vector $\vmx$ in the convex hull of a
finite set $U \subset \R^n$, how many elements of $U$ are needed to approximate
$\vmx$ within a desired level of accuracy? Maurey's empirical method answers
this question by constructing a sequence of random vectors and estimating the
number of elements required for the expected average to fall below a specific
distance to $\vmx$.
Unfortunately, unless the number of groups in the partition $\partI$ is
identical to the ambient dimension $\dimxtot$, the group $\ell_1$ unit ball can
not be expressed as the convex hull of a finite set.\footnote{
  For instance, the group $\ell_1$-ball in $\R^2$ for $\numG = 1$ (and
  therefore $\dimg = 2$) corresponds to the $\ell_2$-ball $\ball{2}^2$.
}
We will circumvent this problem by an additional covering argument.}

{Let $\vmx \in \ball{\partI,1}^\dimxtot$ such that $\sum_{i=1}^\numG
\norm{\vmx_{\partI_i}}_2 \leq 1$, and denote by $S \subset [\numG]$ the index
set of nonzero groups of $\vmx$.
Then we can express $\vmx$ as
\Equation{\label{eqn:vector_sphere_decomposition}
  \vmx
  = \sum_{j \in S} \vmx_{\partI_j} = \sum_{j \in S} \norm{\vmx_{\partI_j}}_2
  \underbrace{\frac{\vmx_{\partI_j}}{\norm{\vmx_{\partI_j}}_2}}_{\in
  \sphere[\partI_j]{\dimxtot-1}}
}
where $\sphere[\partI_j]{\dimxtot-1}$ denotes the subset of the complex unit
sphere in $\C^\dimxtot$ supported on an index set $\partI_j$.
Since Maurey's lemma is concerned with
the estimation of the covering number of the convex hull of a finite point
cloud \wrt an arbitrary metric, the argument does not immediately extend to
the current setting.
This is due to fact for every $\vmx \in \ball{\partI,1}^\dimxtot$, the
dictionary
\Equation*{
  U_\vmx
  \defeq \set{\frac{\vmx_{\partI_j}}{\norm{\vmx_{\partI_j}}_2}}[j \in S]
}
such that $\vmx \in \conv(U_\vmx)$ depends on the particular choice of $\vmx$.
In other words, since $\ball{\partI,1}^\dimxtot$ does not generally admit a
polytope representation, there exists no finite set $U \subset \C^\dimxtot$
such that $\ball{\partI,1}^\dimxtot = \conv(U)$. }

{To deal with the issue outlined above, we establish the following result, which
generalizes Maurey's lemma to more complicated sets.
With some abuse of notation, we first introduce the following generalization of
the convex hull of a set.
Let $\set{\Uset_i}_{i=1}^B$ be a collection of compact subsets in a normed
space.
Then we denote by $\conv_B(\Uset_1, \ldots, \Uset_B)$ the set of convex
combinations with each $\Uset_i$ contributing exactly one element to each
vector $\vmx \in \conv_B(\Uset_1, \ldots, \Uset_B)$.
More precisely, we set
\Equation*{
  \conv_B(\Uset_1,\ldots,\Uset_B) \defeq
  \set{\sum_{i=1}^B\alpha_i\vmu_i}[\sum_{i=1}^B \alpha_i=1, \alpha_i\geq 0,
  \vmu_i \in \Uset_i \Forall i \in [B]]
}
where we use the index $B$ in the notation $\conv_B$ to emphasize the fact that
each element of $\conv_B(\Uset_1, \ldots, \Uset_B)$ consists of exactly $B$
vectors drawn from a different set $\Uset_i$.
If $\Uset \subset \R^\dimxtot$ is a compact subset, then by the Carath\'eodory
theorem, we recover the usual notion of the convex hull of $\Uset$ as
\Equation*{
  \conv(\Uset)
  = \conv_{\dimxtot+1}(\underbrace{\Uset, \ldots,
  \Uset}_{\substack{\text{$\dimxtot+1$} \\ \text{copies of $\Uset$}}}).
}
We point out that the result below also holds if we assume the sets $\Uset_i$
to be both compact and convex in which case we may replace $\conv_B(\Uset_1,
\ldots, \Uset_B)$ by $\conv(\union_{i=1}^B \Uset_i)$.}

{
\Proposition[Maurey's extended lemma]{\label{prop:maurey_extended_lemma}
  Let $(X, \norm{\cdot}_X)$ be a normed space, and let $\Uset_1,\dots, \Uset_B
  \subset X$ be compact sets.
  Assume that for every $K \in \N$ and $\vmz_i \in \union_{j=1}^B \Uset_j$
  with $i = 1,\ldots,K$ the following holds:
  \Equation*{
    \E\norm{\sum_{i=1}^K \epsilon_i \vmz_i}_X
    \leq A \sqrt{K}
  }
  where $(\epsilon_i)_{i=1}^K$ is an independent Rademacher sequence, and
  $A > 0$ is a constant.
  Then for every $u > 0$,
  \Equation*{
    \log\covnum(\conv_B(\Uset_1,\dots,\Uset_B),\norm{\cdot}_X,u)
    \lesssim (A/u)^2 \log\parens{\sum_{i=1}^B
    \covnum(\Uset_i,\norm{\cdot}_X,u/2)}.
  }
}
\Proof{
  We first equip each set $\Uset_i$ with its own net $\net_i$ with covering
  radius $u/2$ \wrt the canonical metric induced by $\norm{\cdot}_X$.
  Next, denote by
  \Equation*{
    \function{\pi_i}{X}{\net_i}[\vmx][\argmin*{\vmz \in \net_i}{\norm{\vmx -
    \vmz}_X}]
  }
  the projection on $\net_i$ in terms of $\norm{\cdot}_X$, and set for $\vmx
  \in X$,
  \Equation*{
    \pi(\vmx)
    \defeq \argmin*{\vmx_0 \in \set{\pi_i(\vmx)}[i \in [B]]}{\norm{\vmx -
    \vmx_0}_X} \in \union_{i=1}^B \net_i.
  }
  Consider now a vector $\vmx \in \conv_B(\Uset_1, \ldots, \Uset_B)$ such that
  \Equation*{
   \vmx
   = \alpha_1 \vmu_1 + \ldots + \alpha_B \vmu_B
  }
  with $\vmu_i \in \Uset_i$ and $\alpha_i \in [0,1]$ for $i \in [B]$ with
  $\sum_{i=1}^B \alpha_i = 1$.
  Since the convex multipliers $\set{\alpha_i}_i$ define a discrete probability
  distribution on $[B]$, this allows us to construct a random vector $\vmz \in
  X$ with
  \Equation*{
    \P(\vmz = \vmu_i) = \alpha_i,
  }
  such that $\E\vmz = \vmx$.
  Consider now $K$ independent copies $\vmz_1, \ldots, \vmz_K$ of $\vmz$.
  Then we have by the triangle inequality that
  \Equation{
    \E \norm{\vmx - \frac{1}{K} \sum_{i=1}^K \pi(\vmz_i)}_X
    &\leq \E \norm{\vmx - \frac{1}{K} \sum_{i=1}^K \vmz_i}_X + \E
    \norm{\frac{1}{K} \sum_{i=1}^K (\vmz_i - \pi(\vmz_i))}_X \notag \\
    &\leq \E \norm{\vmx - \frac{1}{K} \sum_{i=1}^K \vmz_i}_X +
    \frac{1}{K} \sum_{i=1}^K \E\norm{\vmz_i - \pi(\vmz_i)}_X.
    \label{eqn:maurey_triangle_inequality}
  }
  For the summands of the second term we find
  \Equation*{
    \E \norm{\vmz_i - \pi(\vmz_i)}_X
    = \sum_{j=1}^B \alpha_j \norm{\vmu_j - \pi(\vmu_j)}_X
    \leq \sum_{j=1}^B \alpha_j u/2
    = u/2
  }
  since $\pi$ maps every vector $\vmu_j \in \Uset_j$ to its respective
  $(u/2)$-net $\net_j$.
  Next, we focus on the first term in \blaref*{eqn:maurey_triangle_inequality}
  for which we find
  \Equation*{
    \E \norm{\vmx - \frac{1}{K} \sum_{i=1}^K \vmz_i}_X
    = \frac{1}{K} \E \norm{\sum_{i=1}^K (\vmz_i - \E\vmz_i)}_X
  }
  since $\E\vmz_i = \vmx$ for all $\vmz_i$.
  Fixing randomness by conditioning on $\set{\vmz_i}_i \subset \union_{j=1}^B
  \Uset_j$ and invoking the Gin\'e-Zinn symmetrization principle
  \cite{gine1984} then yields
  \Equation*{
    \E\norm{\vmx - \frac{1}{K} \sum_{i=1}^K \vmz_i}_X
    \leq \frac{2}{K} \E \norm{\sum_{i=1}^K \epsilon_i \vmz_i}_X
    \leq \frac{2}{K} A \sqrt{K}
    = \frac{2A}{\sqrt{K}}
  }
  where $(\epsilon_i)_i$ is an independent Rademacher sequence, and the
  last step follows by the assumption of \blaref{prop:maurey_extended_lemma}.
  We therefore find by collecting our estimates that
  \Equation*{
    \E \norm{\vmx - \frac{1}{K} \sum_{i=1}^K \vmz_i}_X
    = \frac{2A}{\sqrt{K}} + \frac{u}{2},
  }
  which implies for
  \Equation*{
    K
    \geq 16 \frac{A^2}{u^2}
  }
  that there exists at least one realization of the random vector
  \Equation*{
    \zhat
    \defeq \frac{1}{K} \sum_{i=1}^K \pi(\vmz_i)
  }
  such that $\norm{\vmx - \zhat}_X \leq u$.
  To complete the proof, it remains to count the number of possible
  realizations of $\zhat$.
  Choosing the nets $\net_i$ as the smallest $(u/2)$-nets, we have
  $\card{\net_i} = \covnum(\Uset_i, \norm{\cdot}_X, u/2)$.
  Since $\pi$ maps any element of $X$ on one of the $B$ nets $\net_i$,
  there are exactly
  \Equation*{
    \parens{\sum_{i=1}^B \covnum(\Uset_i, \norm{\cdot}_X, u/2)}^K
  }
  realizations of $\zhat$.
  Since the above argument holds for any $\vmx \in \conv_B(\Uset_1, \ldots,
  \Uset_B)$, we conclude that
  \Equation*{
    \log\covnum(\conv_B(\Uset_1, \ldots, \Uset_B), \norm{\cdot}_X, u)
    \leq 16 \frac{A^2}{u^2} \log\parens{\sum_{i=1}^B \covnum(\Uset_i,
    \norm{\cdot}_X, u/2)}
  }
  as claimed.
}
}

{From our previous discussion, we have that every vector $\vmx \in
\ball{\partI,1}^\dimxtot$ can be decomposed for $S = \bsupp(\vmx) = \smallset{i
\in [\numG]}[\vmx_{\partI_i} \neq \vmzero]$ as
\Equation*{
  \vmx
  = \sum_{i \in S} \norm{\vmx_{\partI_i}}_2
  \frac{\vmx_{\partI_i}}{\norm{\vmx_{\partI_i}}_2}
}
where each vector $\vmu_i \defeq \vmx_{\partI_i} / \norm*{\vmx_{\partI_i}}_2$
is $1$-group-sparse \wrt the group partition $\partI$ with $\norm{\vmu_i}_2 =
1$ and therefore $\vmu_i \in \sphere[\partI_i]{\dimxtot-1}$.
Note, however, that the choice $\Uset_i = \sphere[\partI_i]{\dimxtot-1}$ in
\blaref{prop:maurey_extended_lemma} does not work since for points $\vmx \in
\interior(\ball{\partI,1}^\dimxtot)$, we have
\Equation*{
  \sum_{i=1}^\numG \norm{\vmx_{\partI_i}}_2
  = \sum_{i=1}^\numG \alpha_i
  = \bnorm{\vmx}{1} < 1
}
and hence $\vmx \notin \conv_\numG(\sphere[\partI_1]{\dimxtot-1}, \ldots,
\sphere[\partI_\numG]{\dimxtot-1})$ since the definition of
$\conv_\numG(\sphere[\partI_1]{\dimxtot-1}, \ldots,
\sphere[\partI_\numG]{\dimxtot-1})$ assumes that its elements consist of convex
combinations of exactly $\numG$ elements with $\sum_{i=1}^\numG \alpha_i = 1$.
Instead, we may either choose $\Uset_i = \sphere[\partI_i]{\dimxtot-1} \cup
\set{\vmzero}$ or $\Uset_i = \ball{2}^\dimxtot \cap \C_{\partI_i}^\dimxtot =
\ball{\set{\partI_i},1}^\dimxtot$.
We choose the latter option here since the volume comparison argument we will
use below to bound the covering number of each $\Uset_i$ (\wrt
$\norm{\cdot}_2$) yields the same bound for both $\sphere[\partI_i]{\dimxtot-1}
\cup \set{\vmzero}$ and $\ball{2}^\dimxtot \cap \C_{\partI_i}^\dimxtot$ since
$\sphere[\partI_i]{\dimxtot-1} \cup \set{\vmzero} \subset \ball{2}^\dimxtot
\cap \C_{\partI_i}^\dimxtot$.}

{
\Proposition{\label{prop:group_l1_ball_covnum}
  For the covering number of the group $\ell_1$ unit ball \wrt the
  canonical metric induced by $\norm{\cdot}_V$ it holds that
  \Equation*{
    \sqrt{\log\covnum(\ball{\partI,1}^\dimxtot, \norm{\cdot}_V, u)}
    \lesssim u^{-1} \coherence(\vmPsi) \sqrt{\log(\dimxtot)}
    \parens{\sqrt{\log(\numG)} + \sqrt{2\dimg \log(1 + 4 / u)}}.
  }
}
\Proof{
  As discussed above, we choose $\Uset_i = \ball{2}^\dimxtot \cap
  \C_{\partI_i}^\dimxtot$ in \blaref{prop:maurey_extended_lemma} and
  equip each unit ball in the coordinate subspace $\C_{\partI_i}^\dimxtot$
  supported on $\partI_i$ with a net $\net_i$ of covering radius $u/2$.
  Given a vector $\vmx \in \ball{\partI,1}^\dimxtot = \conv_\numG(\Uset_1,
  \ldots, \Uset_\numG)$, it merely remains to find an appropriate bound on the
  expected norm of the Rademacher sum $\E \norm*{\sum_{i=1}^K \epsilon_i
  \vmz_i}_V$ for $K$ vectors $\vmz_1, \ldots, \vmz_K \in \union_{i=1}^\numG
  \Uset_i$.
  To that end, first note that we have by the definition of $\norm{\cdot}_V =
  \opnorm{V(\cdot)}$ and linearity of the operator $V$ (\cf
  \blaref{eqn:V_operator}) that
  \Equation*{
    \E \norm{\sum_{i=1}^K \epsilon_i \vmz_i}_V
    = \E \opnorm{\sum_{i=1}^K \epsilon_i V(\vmz_i)}.
  }
  Next, we invoke the following noncommutative Khintchine inequality for
  operator norms due to Eftekhari \etal}
  }

{
  \Lemma[\cite[\lemName~9]{Eftekhari2012TheRI}]{\label{lem:khintchine_inequality}
    Let $\set{\vmV_i}_{i=1}^K$ be a collection of matrices with the same
    dimension and rank at most $r$.
    Denote by $(\epsilon_i)_{i=1}^K$ an independent Rademacher sequence.
    Then
    \Equation*{
      \E \opnorm{\sum_{i=1}^K \epsilon_i \vmV_i}
      \lesssim \sqrt{\log(r)} \parens{\sum_{i=1}^K \opnorm{\vmV_i}^2}^{1/2}.
    }
  }
}

{
Since the operator $V$ yields for any $\vmx \in \C^\dimxtot$ a matrix of
  size $\dimy\numO \times \dimy\dimx\numO$, we have $\rank V(\vmz_i) \leq
  \dimy\numO = \dimytot$.
  An application of \blaref{lem:khintchine_inequality} therefore yields
  \Equation*{
    \E \norm{\sum_{i=1}^K \epsilon_i \vmz_i}_V
    &\lesssim \sqrt{\log(\dimytot)} \parens{\sum_{i=1}^K
    \opnorm{V(\vmz_i)}^2}^{1/2} \\
    &\leq \sqrt{\log(\dimytot)} \parens{\sum_{i=1}^K \coherence(\vmPsi)^2
    \bnorm{\vmz_i}{1}^2}^{1/2} \\
    &\leq \coherence(\vmPsi) \sqrt{\log(\dimytot)} \sqrt{K}
  }
  where the second step is due to \blaref*{eqn:induced_norm_simple_bound}, and the last
  step follows since each vector $\vmz_i \in \Uset_i$ is $1$-group-sparse \wrt
  $\partI$ by construction.}

{
  To complete the proof, we need to bound the covering numbers of the
  coordinate-restricted unit balls $\ball{2}^\dimxtot \cap
  \C_{\partI_i}^\dimxtot$.
  Assuming that we have for each net $\net_i$ that $\card{\net_i} =
  \covnum(\ball{2}^\dimxtot \cap \C_{\partI_i}^\dimxtot, \norm{\cdot}_V, u/2)$,
  we denote by $\nu \defeq \Max[i \in [\numG]]{\card{\net_i}}$ the cardinality
  of the biggest net.
  By the volume comparison argument for norm balls 
  we find with
  \blaref*{eqn:induced_norm_simple_bound} that
  \Equation*{
    \card{\net_i}
    &= \covnum(\ball{2}^\dimxtot \cap \C_{\partI_i}^\dimxtot, \norm{\cdot}_V,
    u/2) \\
    &\leq \covnum(\ball{2}^{\dimg_i}, \norm{\cdot}_2, u/2) \\
    &\leq \parens{1 + \frac{2}{u/2}}^{2\dimg_i}
  }
  and therefore
  \Equation*{
    \nu \leq \parens{1 + \frac{4}{u}}^{2\dimg}
  }
  with $\dimg = \Max[i \in [\numG]]{\dimg_i}$ as usual.
  The factor $2$ in the exponent is again due to isomorphic identification of
  $\C^{\dimg_i}$ with $\R^{2\dimg_i}$.
  Combining this estimate with $A \lesssim
  \coherence(\vmPsi)\sqrt{\log(\dimytot)} \leq
  \coherence(\vmPsi)\sqrt{\log(\dimxtot)}$, we finally find by invoking
  \blaref{prop:maurey_extended_lemma} that
  \Equation{
    \sqrt{\log\covnum(\ball{\partI,1}^\dimxtot, \norm{\cdot}_V, u)}
    &= \sqrt{\log\covnum(\conv_\numG(\ball{2}^\dimxtot \cap
    \C_{\partI_1}^\dimxtot, \ldots, \ball{2}^\dimxtot \cap
    \C_{\partI_\numG}^\dimxtot), \norm{\cdot}_V, u)} \notag \\
    &\lesssim \frac{A}{u} \sqrt{\log\parens{\sum_{i=1}^\numG \card{\net_i}}}
    \notag \\
    &\leq \frac{\coherence(\vmPsi) \sqrt{\log(\dimxtot)}}{u}
    \parens{\sqrt{\log(\numG\nu)}} \notag \\
    &\leq \frac{\coherence(\vmPsi) \sqrt{\log(\dimxtot)}}{u}
    \parens{\sqrt{\log(\numG)} + \sqrt{2\dimg\log\parens{1 +
    \frac{4}{u}}}}. \label{eqn:covnum_bound_2}
  }
  This completes the proof.
}

To establish our final bound on the $\gamFunc$-functional of $\opimg$, we split
the entropy integral in two parts according to \blaref*{eqn:entropy_integral_parts}.
We then control the first part via the
volume comparison estimate \blaref*{eqn:covnum_bound_1} and bound the second
integral via \blaref{prop:group_l1_ball_covnum}.
For the first integral, this yields\footnote{
  The last estimate follows from the bound $\int_0^\alpha \sqrt{\log(1 +
  t^{-1})} \de{t} \leq \alpha \sqrt{\log(e(1 + \alpha^{-1}))}$ for $\alpha > 0$
  (see, \eg, \cite[\lemName~C.9]{Foucart2013mathematicalIC}).
}
\Equation{
  \int_0^\lambda \sqrt{\log\covnum(\sigsetO, \norm{\cdot}_V,
  \varepsilon)} \de{\varepsilon}
  &\leq \int_0^\lambda \sqrt{\dims\log(e\numG/\dims) +
  2\dims\dimg\log(1+4/\varepsilon)} \de{\varepsilon} \notag \\
  &\leq \lambda\sqrt{\dims\log(e\numG/\dims)} +
  \lambda\sqrt{2\dims\dimg\log(5e/\lambda)} \label{eqn:entropy_integral_1}.
}
For the second integral in \blaref*{eqn:entropy_integral_parts}, 
note that we have by the Cauchy-Schwarz inequality that
\Equation*{
  \frac{\sigsetO}{\sqrt{\dims}} \subset \ball{\partI,1}^\dimxtot.
}
By a change of variable, this yields 
\Equation*{
  \int_\lambda^{\radOp(\opimg)} \sqrt{\log\covnum(\sigsetO, \norm{\cdot}_V, u)}
  \de{u}
  &\leq \int_\lambda^{\sqrt{\dims}\coherence(\vmPsi)}
  \sqrt{\log\covnum\parens{\frac{\sigsetO}{\sqrt{\dims}}, \norm{\cdot}_V,
  \frac{u}{\sqrt{\dims}}}} \de{u} \\
  &\leq \sqrt{\dims} \int_{\lambda/\sqrt{\dims}}^{\coherence(\vmPsi)}
  \sqrt{\log\covnum\parens{\ball{\partI,1}^\dimxtot, \norm{\cdot}_V, \frac{u}{2}}} \de{u}.
}
With \blaref{prop:group_l1_ball_covnum} we then find with
\Equation*{
  &\Phantom{\lesssim}\int_\lambda^{\sqrt{\dims}\coherence(\vmPsi)}
  \sqrt{\log\covnum(\sigsetO, \norm{\cdot}_V, \varepsilon)} \de{\varepsilon} \\
  &\lesssim 2\sqrt{\dims}\coherence(\vmPsi)\sqrt{\log(\dimxtot)}
  \parens{\int_{\lambda / (2\sqrt{\dims})}^{\coherence(\vmPsi)/2}
  \varepsilon^{-1} \sqrt{\log(\numG)} \de{\varepsilon} + \int_{\lambda /
  (2\sqrt{\dims})}^{\coherence(\vmPsi)/2} \varepsilon^{-1} \sqrt{\dimg\log(1 +
  8/\varepsilon)} \de{\varepsilon}}
}
For the last integral, note that $\sqrt{\log(1+t^{-1})}$ is monotonically
decreasing in $t$.
Hence, we have that
\Equation*{
  \int_a^b t^{-1} \sqrt{\log(1+t^{-1})} \de{t}
  \leq \log(b/a) \sqrt{\log(1+a^{-1})}.
}
This yields
\Equation{
  &\Phantom{\lesssim} \int_\lambda^{\sqrt{\dims}\coherence(\vmPsi)}
  \sqrt{\log\covnum(\sigsetO, \norm{\cdot}_V, \varepsilon)} \de{\varepsilon}
  \notag \\
  &\lesssim \sqrt{\dims} \coherence(\vmPsi) \sqrt{\log(\dimxtot)}
  \log(\sqrt{\dims}\coherence(\vmPsi) / \lambda)
  \parens{\sqrt{\log(\numG)} + \sqrt{\dimg\log(1 + 16\sqrt{\dims} / \lambda)}}.
  \label{eqn:entropy_integral_2}
}
Compared with our previous estimate based on Sudakov's inequality for which we
found
\Equation*{
  \int_\lambda^{\sqrt{\dims}\coherence(\vmPsi)}
  \sqrt{\log\covnum(\sigsetO, \norm{\cdot}_V, \varepsilon)} \de{\varepsilon}
  \lesssim \sqrt{\dims}
  \log(\sqrt{\dims}\coherence(\vmPsi) / \lambda)
  \parens{\sqrt{\log(\numG)} + \sqrt{\dimg}},
}
our new bound differs by an additional $\log$-factor in $\dimxtot$, as well as
another logarithmic factor depending on $\lambda$.
However, we also obtain the desired linear dependence on $\coherence(\vmPsi)$.
Simplifying \blaref*{eqn:entropy_integral_1} and
\blaref*{eqn:entropy_integral_2} by absorbing numerical constants into the
implicit constant in the notation and collecting both estimates, we eventually
find
\Equation*{
  \gamFunc(\opimg, \opnorm{\cdot})
  &\lesssim \lambda\sqrt{\dims\log(\numG/\dims)} +
  \lambda\sqrt{\dims\dimg\log(1/\lambda)} \\
  &\quad + \sqrt{\dims}\coherence(\vmPsi)\sqrt{\log(\dimxtot)}
  \log(\dims\coherence(\vmPsi) / \lambda) \parens{\sqrt{\log(\numG)} +
  \sqrt{\dimg\log(\dims / \lambda)}},
}
which, for the choice $\lambda = \coherence(\vmPsi)$, ultimately
results in
\Equation{
  \gamFunc(\opimg, \opnorm{\cdot})
  &\lesssim \coherence(\vmPsi)\sqrt{\dims\log(\numG/\dims)} +
  \coherence(\vmPsi)\sqrt{\dims\dimg\log(1/\coherence(\vmPsi))} \notag \\
  &\quad + \sqrt{\dims}\coherence(\vmPsi)\sqrt{\log(\dimxtot)}
  \log(\dims) \parens{\sqrt{\log(\numG)} +
  \sqrt{\dimg\log(\dims/\coherence(\vmPsi))}} \notag \\
  &\lesssim \sqrt{\dims}\coherence(\vmPsi)\sqrt{\log(\dimxtot)} \log(\dims)
  \parens{\sqrt{\log(\numG)} + \sqrt{\dimg\log(\dims/\coherence(\vmPsi))}}.
  \label{eqn:gamma_2_final_bound}
}

To establish our main result, it remains to invoke \blaref{thm:chaos_bound}
after collecting our estimates for $\radFrob(\opimg), \radOp(\opimg)$ and
$\gamFunc(\opimg, \opnorm{\cdot})$.
This concludes the proof of \blaref{thm:block_diagonal_group_rip}.

\section{The \texorpdfstring{\Acs{GRIP}}{Group-RIP} for Block Diagonal Matrices
with Constant Block-Diagonal}
\label{sec:shared_sensing_matrix}

Let us now turn to the scenario in which each sensor is equipped with a
copy of the same measurement matrix $\vmPhi$, \ie, we observe
\Equation*{
  \vmy = \vmA\vmPsi\vmx = \PMatrix{
    \vmPhi & & \\
    & \ddots & \\
    & & \vmPhi
  } \vmPsi \vmx =
  \PMatrix{\vmPhi\vmPsi_1\vmx \\ \vdots \\ \vmPhi\vmPsi_\numO\vmx}.
}
While we could use the same transformations $V_l$ as in the case of unique
per-sensor matrices, and set
\Equation*{
  \vmA\vmPsi\vmx \eqdist \PMatrix{V_1(\vmx) \\ \vdots \\ V_\numO(\vmx)} \vmxi
  =: V'(\vmx) \vmxi
}
with $\vmxi \in \R^{\dimy\dimx}$ a unit-variance $\tau$-subgaussian random
vector, the lack of a (block) diagonal structure in $V'$ complicates the
calculation of both $\radOp$ and $\gamFunc$ as we cannot concisely express the
operator norm in terms of a mixed $(\ell_\infty,\ell_2)$ vector norm as in
\blaref{eqn:induced_norm}. However, since we only require
$\norm{\vmA\vmPsi\vmx}_2^2$ and $\norm{V'(\vmx)\vmxi}_2^2$ to be identical in
distribution to apply \blaref{thm:chaos_bound}, we are free to reorder the rows
of $V'(\vmx)$. To that end, we define the operator
\Equation*{
  \Vtilde(\vmx) = \PMatrix{
    \transp{(\vmPsi_1\vmx)} \\
    \vdots \\
    \transp{(\vmPsi_\numO\vmx)}
  } \in \C^{\numO \times \dimx}.
}
Then we have with the block diagonal matrix
\Equation*{
  \Vhat(\vmx) := \PMatrix{
    \Vtilde(\vmx) & & \\
    & \ddots & \\
    & & \Vtilde(\vmx)
  } \in \C^{\dimy\numO \times \dimy\dimx}
}
with $\dimy$ copies of $\Vtilde(\vmx)$ on its diagonal that
$\norm{\vmA\vmPsi\vmx}_2^2 \eqdist \norm*{\Vhat(\vmx)\vmxi}_2^2$. As before, we
define the set $\opimghat = \Vhat(\sigsetO)$ so that
\Equation*{
  &\Phantom{=}\P\parens{\sup_{\vmx \in \sigsetO}
  \abs{\norm{\frac{1}{\sqrt{\dimy}}\vmA\vmPsi\vmx}_2^2 - 1} \geq \delta}
  = \P\parens{\frac{1}{\dimy} \sup_{\vmGamma \in \opimghat}
  \abs{\norm{\vmGamma\vmxi}_2^2 - \E\norm{\vmGamma\vmxi}_2^2} \geq \delta}.
}
It remains to estimate the radii of $\opimghat$, as well as its metric
entropy integral. Unsurprisingly, we mostly proceed in the same way as before.
For convenience of notation, we associate with $\Vhat$ the norm
$\norm{\cdot}_\Vhat$ on $\C^\dimxtot$ induced by $\norm{\cdot}_\Vhat =
\opnorm*{\Vhat(\cdot)}$.

First, note that
\Equation*{
  \Fnorm{\Vhat(\vmx)}^2 &= \sum_{i=1}^\dimy \Fnorm{\Vtilde(\vmx)}^2 =
  \dimy \sum_{l=1}^\numO \tr(\Vtilde(\vmx)\adj{\Vtilde(\vmx)}) \\
  &= \dimy \sum_{l=1}^\numO \norm{\vmPsi_l\vmx}_2^2 = \dimy \norm{\vmx}_2^2,
}
and therefore
\Equation*{
  \radFrob(\opimghat) = \sup_{\vmGamma \in \opimghat}
  \Fnorm{\vmGamma} = \sup_{\vmx \in \sigsetO} \Fnorm{\Vhat(\vmx)} = \sup_{\vmx
  \in \sigsetO} \sqrt{\dimy} \norm{\vmx}_2 = \sqrt{\dimy}.
}

Next, denote as before by $S \subset [\numG]$ the index set of nonzero groups
of $\vmx \in \C^{\dimxtot}$ \wrt $\partI$. Then we have due to linearity of
$\Vtilde$, and consequently linearity of $\Vhat$ that
\Equation{
  \opnorm{\Vhat(\vmx)} &= \opnorm{\Vtilde(\vmx)} =
  \opnorm{\sum_{i=1}^\numG \Vtilde\parens{\vmx_{\partI_i}}} \notag \\
  &\leq \sum_{i \in S} \norm{\vmx_{\partI_i}}_2 \cdot
  \opnorm{\Vtilde\parens{\frac{\vmx_{\partI_i}}{\norm{\vmx_{\partI_i}}_2}}}
  \notag \\
  &\leq \bnorm{\vmx}{1} \max_{i \in S}{
    \opnorm{\Vtilde\parens{\frac{\vmx_{\partI_i}}{\norm{\vmx_{\partI_i}}_2}}}}
    \notag \\
  &\leq \bnorm{\vmx}{1} \max_{i \in [\numG]}{\sup_{\vmu \in
  \sphere[\partI_i]{\dimxtot-1}} \opnorm{\Vtilde(\vmu)}}
  \label{eqn:vhat_estimate}.
}
In the edge case where the number of groups $\numG$ coincides with the ambient
dimension $\dimxtot$ (\ie, in case of regular sparsity rather than
group-sparsity), the supremum in \blaref{eqn:vhat_estimate} can be easily
computed as each sphere $\sphere[\partI_i]{\dimxtot-1}$ reduces \Wlog to a
two-element\footnote{In light of the linearity of $\Vtilde$, this in turn
implies the supremum in \blaref*{eqn:vhat_estimate} is taken over a singleton
set.} set $\set{\pm\vme^i}$ where $\vme^i \in \R^\dimxtot$ denotes the $i$-th
canonical unit vector. However, the same does not hold for $\numG < \dimxtot$
which does not allow us to compute \blaref*{eqn:vhat_estimate} numerically. To
circumvent the computability issue, we estimate the supremum as follows.

Denote by $\vmu$ an arbitrary unit-normalized $1$-group-sparse vector \wrt the
group partition $\partI$. Then
\Equation*{
  &\Phantom{=} \opnorm{\Vtilde(\vmu)} \\
  &= \sup_{\vmz \in \ball{2}^\dimx}
  \norm{\Vtilde(\vmu)\vmz}_2 \leq \sqrt{\numO} \sup_{\vmz \in \ball{2}^\dimx}
  \norm{\Vtilde(\vmu)\vmz}_\infty \\
  &= \sqrt{\numO} \sup_{\vmz \in
  \ball{2}^\dimx} \max_{l \in [\numO]} \abs{\inner{\vmPsi_l\vmu}{\vmz}}
  = \sqrt{\numO} \sup_{\vmz \in \ball{2}^\dimx} \max_{l \in [\numO]}
  \abs{\inner{\vmu}{\transp{\vmPsi_l}\vmz}} \\
  &\leq \sqrt{\numO} \bnorm{\vmu}{1}
  \sup_{\vmz \in \ball{2}^\dimx} \max_{l \in [\numO]}
  \bnorm{\transp{\vmPsi_l}\vmz}{\infty} \\
  &= \sqrt{\numO} \max_{l \in [\numO]} \sup_{\vmz \in \ball{2}^\dimx}
  \bnorm{\transp{\vmPsi_l}\vmz}{\infty}
}
where we used the fact that $\bnorm{\vmu}{1} = 1$ since $\vmu$ is a unit-norm
vector supported on a single group in $\partI$. Expanding the supremum, we find
\Equation*{
  &\Phantom{=} \sup_{\vmz \in \ball{2}^\dimx}
  \bnorm{\transp{\vmPsi_l}\vmz}{\infty} \\
  &= \sup_{\vmz \in \ball{2}^\dimx} \max_{i \in [\numG]}
  \norm{(\transp{\vmPsi_l} \vmz)_{\partI_i}}_2
  = \max_{i \in [\numG]} \sup_{\vmz \in \ball{2}^{\card*{\partI_i}}}
  \norm{\transp{\parens{(\vmPsi_l)_{\partI_i}}}\vmz}_2 \\
  &= \max_{i \in [\numG]} \opnorm{\transp{\parens{(\vmPsi_l)_{\partI_i}}}} =
  \max_{i \in [\numG]} \opnorm{(\vmPsi_l)_{\partI_i}},
}
where $(\vmPsi_l)_{\partI_i} \in \C^{\dimx \times \card{\partI_i}}$ denotes the
submatrix of $\vmPsi_l$ restricted to the columns indexed by $\partI_i$. The two
estimates therefore yield
\Equation{\label{eqn:vhat_bnorm}
  \opnorm{\Vhat(\vmx)} \leq \bnorm{\vmx}{1} \sqrt{\numO} \max_{\substack{i \in
  [\numG], \\ l \in [\numO]}}{\opnorm{(\vmPsi_l)_{\partI_i}}}.
}
Unfortunately, this bound is too loose in the previously discussed edge case
where $\numG = \dimxtot$ as it does not reduce to
\Equation*{
  \opnorm{\Vhat(\vmx)} \leq \norm{\vmx}_1 \max_{i \in
  [\dimxtot]}{\opnorm{\Vtilde(\vme^i)}}
}
which immediately follows from \blaref{eqn:vhat_estimate}. In other words, the
bound does not reduce to the natural bound we obtain in the sparse setting. To
remedy the situation, we also consider the following bound. Note that for $i
\in S = \set{i \in [\numG]}[\vmx_{\partI_i} \neq \vmzero]$, we have
\Equation*{
  &\Phantom{\leq}
  \opnorm{\Vtilde\parens{\frac{\vmx_{\partI_i}}{\norm{\vmx_{\partI_i}}_2}}} \\
  &\leq \sup_{\vmu \in \sphere[\partI_i]{\dimxtot-1}} \opnorm{\Vtilde(\vmu)}
  = \sup_{\vmu \in \sphere[\partI_i]{\dimxtot-1}} \opnorm{\sum_{j \in \partI_i}
  \abs{u_j} \Vtilde(\vme^j)} \\
  &\leq \sup_{\vmu \in \sphere[\partI_i]{\dimxtot-1}} \sum_{j \in \partI_i}
  \abs{u_j} \cdot \opnorm{\Vtilde(\vme^j)} \\
  &\leq \sup_{\vmu \in
  \sphere[\partI_i]{\dimxtot-1}} \norm{\vmu}_1 \max_{j \in
  \partI_i}{\opnorm{\Vtilde(\vme^j)}}
  \leq \sqrt{\card{\partI_i}} \max_{j \in \partI_i}{\opnorm{\Vtilde(\vme^j)}} \\
  &\leq \sqrt{\dimg} \max_{j \in \partI_i}{\opnorm{\Vtilde(\vme^j)}}.
}
Combining both estimates in the parameter
\Equation*{
  &\orthoconst(\vmPsi) = \\
  &\Phantom{=}\min\set{
    \sqrt{\dimg} \max_{i \in [\dimxtot]}{\opnorm{\Vtilde(\vme^i)}},
    \sqrt{\numO} \max_{\substack{l \in [\numO], \\ i \in
    [\numG]}}{\opnorm{(\vmPsi_l)_{\partI_i}}}
  },
}
we find
\Equation*{
  \opnorm{\Vhat(\vmx)} \leq \bnorm{\vmx}{1} \orthoconst(\vmPsi)
}
which finally yields
\Equation*{
  \radOp(\opimghat) &= \sup_{\vmGamma \in \opimghat} \opnorm{\vmGamma}
  = \sup_{\vmx \in \sigsetO} \opnorm{\Vhat(\vmx)} \\
  &\leq \sup_{\vmx \in \sigsetO} \bnorm{\vmx}{1} \orthoconst(\vmPsi)
  \leq \sqrt{\dims} \orthoconst(\vmPsi).
}

Now note that we have
\Equation*{
  \opnorm{\Vhat(\vmx)}^2 &= \opnorm{\Vtilde(\vmx)}^2 =
  \opnorm{\transp{\Vtilde(\vmx)}}^2 \leq \Fnorm{\transp{\Vtilde(\vmx)}}^2 \\
  &= \Fnorm{(\vmPsi_1\vmx, \ldots, \vmPsi_\numO\vmx)}^2 = \sum_{l=1}^\numO
  \norm{\vmPsi_l\vmx}_2^2 \\
  &= \norm{\vmPsi\vmx}_2^2 = \norm{\vmx}_2^2.
}
Estimating the $\gamFunc$-functional of $\opimghat$ by means of the metric
entropy integral
\Equation*{
  \gamFunc(\opimghat, \opnorm{\cdot}) &\lesssim
  \int_0^{\radOp(\opimghat)} \sqrt{\log\covnum(\opimghat,
  \opnorm{\cdot}, \varepsilon)} \de{\varepsilon} \\
  &= \int_0^{\sqrt{\dims}\orthoconst(\vmPsi)} \sqrt{\log\covnum(\sigsetO,
  \norm{\cdot}_\Vhat, \varepsilon)} \de{\varepsilon}
}
therefore proceeds identically to the derivation in
\blaref{sec:metric_entropy_bound}, and thus
\Equation*{
  &\Phantom{\lesssim} \gamFunc(\opimghat, \opnorm{\cdot}) \lesssim
  \sqrt{\dims}\orthoconst(\vmPsi)\sqrt{\log(\dimxtot)} \log(\dims)
  \parens{\sqrt{\log(\numG)}
  \sqrt{\dimg\log(\dims\orthoconst(\vmPsi))}}.
}
In particular, as in the case of \blaref{thm:block_diagonal_group_rip},
\blaref{thm:same_block_diagonal_group_rip} immediately follows by invoking
\blaref{thm:chaos_bound} with the respective estimates of $\radFrob(\opimghat),
\radOp(\opimghat)$ and $\gamFunc(\opimghat, \opnorm{\cdot})$.

\section{{Numerical Results}}
\label{sec:numerical}

{
We now turn to an empirical investigation of the group-sparse recovery problem
from block diagonal observations in terms of the so-called \emph{phase
transition} phenomenon.
Such phenomena collectively describe the sudden change in behavior of a system
when certain parameters cross a critical threshold.
In the \acl{CS} literature, it has been observed early on that such a critical
line exists where recovery of $\dims$-sparse vectors in $\R^\dimx$ from $\dimy$
measurements changes from almost certain success to almost certain failure when
the number of measurements and the sparsity level varies over the half-open
unit square $(\dimy/\dimx, \dims/\dimy) \in (0,1]^2$.
A substantial body of research has since been dedicated to explain, predict and
quantify both the position, as well as the width of the transition region
\cite{donoho2005neighborliness, donoho2006high, donoho2009counting,
donoho2009observed, donoho2010counting, donoho2010exponential}.
The first result to rigorously ascertain the phase transition behavior in the
nonasymptotic regime for Gaussian measurement ensembles was reported by
Amelunxen \etal in \cite{Amelunxen13livingon}.
Their work, which exposes a deep connection between successful recovery via
$\ell_1$-minimization and the concentration behavior of so-called
\emph{intrinsic volumes} in the theory of conic integral geometry, first
managed to not only establish that recovery succeeds in one region, but also
that recovery will fail with high probability in the other.
This is in stark contrast to previous results which were only able to predict
the position of the success region but otherwise could not assess whether
recovery would succeed or fail in the other.}

{
Throughout our experiments, we consider vectors $\xorg \in \C^\dimxtot$ with
$\dimxtot = 1000$.
For a fixed number of $\numO$ sensors, we draw $\numO$ random matrices
$\widetilde{\vmPhi}_l \in \R^{\dimx \times \dimx}$ populated by independent
standard Gaussian random variables.
These matrices are then fixed throughout the process of generating one phase
transition diagram.
Given a pair $(\dimy, \dims)$, we construct the individual sensing matrices
$\vmPhi_l \in \R^{\dimy \times \dimxtot}$ by retaining the first $\dimy$ rows
of each square matrix $\widetilde{\vmPhi}_l$ to form the compound block
diagonal sensing matrix $\vmA = \diag{\dimy^{-1} \vmPhi_l}_{l=1}^\numO$.
We partition the index set $[\dimxtot]$ into $\numG = 100$ nonoverlapping
groups $\partI = \set{\partI_1, \ldots, \partI_{100}}$ such that every group
$\partI_i$ contains $\dimg = 10$ elements.
To that end, we shuffle the elements of the set $[\dimxtot]$ and split them
into $\numG$ groups, which we fix throughout all experiments.
For each of the $50 \times 50$ parameter combinations $(\dimy,\dims)$, we draw
$20$ $\dims$-group-sparse vectors $\xorg \in \C^\dimxtot$, which we recover via
\blaref{prob:group_qcbp}.
Given the group partition $\partI$, we draw the set of active groups uniformly
at random from $[\numG]$.
The nonzero entries in each group are then populated by circularly symmetric
Gaussian random variables.
In other words, given an active group index $k \in S = \bsupp(\xorg)$, we set
$\xorg_{\comp{\partI_k}} = \vmzero$ and $\xorg_{\partI_k} = 2^{-1/2}(\vmg_k + i
\vmh_k)$ where $\vmg_k,\vmh_k \in \R^\dimg$ denote two independent standard
Gaussian random vectors and $i = \sqrt{-1}$.
We then measure how many vectors are successfully recovered according to the
success criterion
\Equation*{
  \frac{\norm{\xorg - \opt{\vmx}}_2}{\norm{\xorg}_2}
  \leq 10^{-3}
}
with $\opt{\vmx}$ denoting the optimal solution of
\blaref{prob:group_qcbp} for $\vmy = \diag{\vmPhi_l}_{l=1}^\numO
\vmPsi \xorg$.
We repeat this experiment for two different sparsity bases $\vmPsi \in
\Ugroup(\dimxtot)$ at the low and high end of the coherence spectrum, namely
the \ac{DFT} and the canonical basis.}

\wraptikz{%
  \begin{figure}[tp]
    \centering
    \begin{subfigure}[b]{0.495\columnwidth}
      \centering
      \includepdf{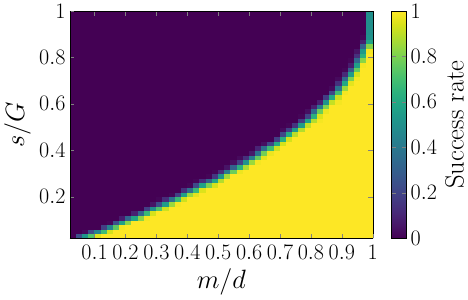}
      \caption{$\numO = 1$}
    \end{subfigure}
    \begin{subfigure}[b]{0.495\columnwidth}
      \centering
      \includepdf{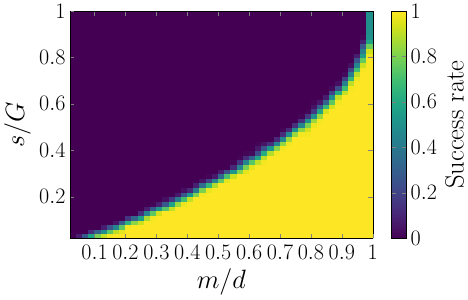}
      \caption{$\numO = 5$}
    \end{subfigure}
    \par\bigskip
    \begin{subfigure}[b]{0.495\columnwidth}
      \centering
      \includepdf{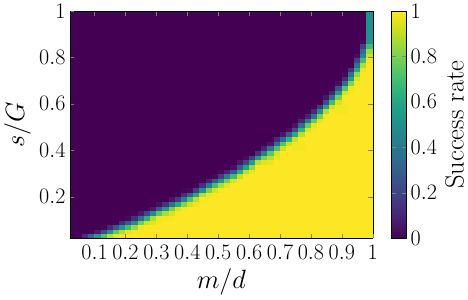}
      \caption{$\numO = 10$}
    \end{subfigure}
    \begin{subfigure}[b]{0.495\columnwidth}
      \centering
      \includepdf{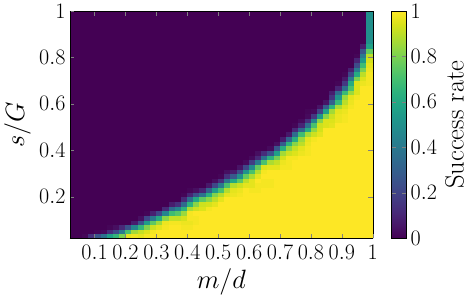}
      \caption{$\numO = 20$}
    \end{subfigure}
    \caption{
      Phase transition diagrams for different numbers of sensors ($\numO$) with
      $\vmPsi = \Id_\dimxtot$ when the group-sparsity level $\dims$ and the
      number of measurements $\dimy$ per sensor vary, and the number of groups
      $\numG$ and the signal dimension per block $\dimx$ is fixed
    }
    \label{fig:group_phase_transition_canonical}
  \end{figure}%
}

{The results of the first set of experiments in which we investigate the
recovery of group-sparse vectors \wrt the canonical basis are shown in
\blaref{fig:group_phase_transition_canonical}.\footnote{
  Note that we normalize abscissa and ordinate by $\dimxtot$ and $\numG$,
  respectively.
  In phase transition diagrams for sparse recovery, it is often more desirable
  to normalize the ordinate by $\dimytot$ to magnify the transition behavior at
  lower values of $\dimytot$.
  This is motivated by the fact that there is no hope to recover an
  $\dims$-sparse vector in $\C^\dimxtot$ from fewer than $\dims$ observations.
  In other words, for a fixed $\dimytot$, it suffices to consider the range
  $\dims \in (0, \dimytot]$.
  In our case, however, this would severely limit resolution since an
  $\dims$-group-sparse vector has $\dims \cdot \dimg$ rather than $\dims$
  nonzero entries.
  By considering $50$ uniformly spaced values for $\dims$, this implies for
  $\dimg = 10$ that the lowest value we can consider on the abscissa would be
  $\dimytot / \dimxtot = \dimy / \dimx = 0.5$.
  Considering that this excludes half the range for $\dimy$, we therefore opt
  to consider the full range of values for $\dims$ between $1$ and $\numG$ for
  every fixed $\dimy$.
}
Despite the fact that our bound does not predict that the number of
measurements required per sensor for $\vmA$ to satisfy the \ac{GRIP} decreases
linearly with $\numO$, the differences in performance are much less dramatic
than one might anticipate.
The biggest differences are observed for small values of $\dimy$.
More precisely, for $\numO = 1$, the transition line tapers off slightly more
for $\dimy \to 0$ compared to the scenario where $\vmA$ contains $\numO = 20$
blocks.
Additionally, it appears that the transition zone where the empirical recovery
rate changes from successful recovery with probability $1$ to $0$ slightly
widens as $\numO$ increases.}

\wraptikz{%
  \begin{figure}[tp]
    \centering
    \begin{subfigure}[b]{0.495\columnwidth}
      \centering
      \includepdf{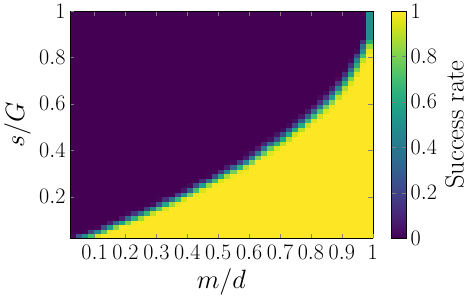}
      \caption{$\numO = 1$}
    \end{subfigure}
    \begin{subfigure}[b]{0.495\columnwidth}
      \centering
      \includepdf{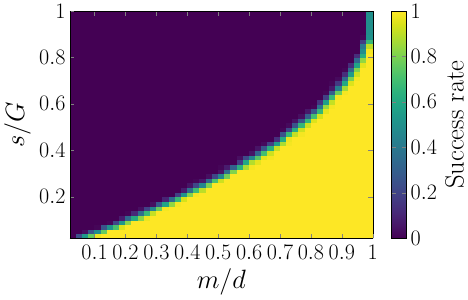}
      \caption{$\numO = 5$}
    \end{subfigure}
    \par\bigskip
    \begin{subfigure}[b]{0.495\columnwidth}
      \centering
      \includepdf{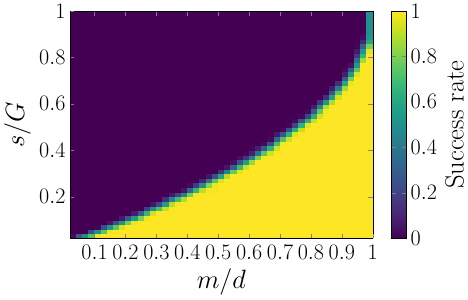}
      \caption{$\numO = 10$}
    \end{subfigure}
    \begin{subfigure}[b]{0.495\columnwidth}
      \centering
      \includepdf{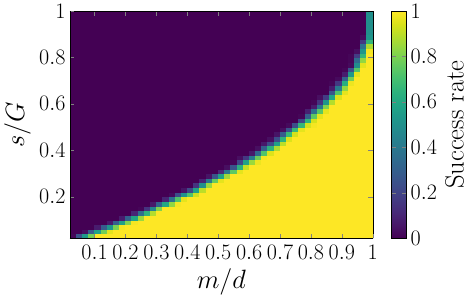}
      \caption{$\numO = 20$}
    \end{subfigure}
    \caption{
      Phase transition diagrams for different numbers of sensors with
      $\vmPsi = \vmF_\dimxtot$
    }
    \label{fig:group_phase_transition_fourier}
  \end{figure}%
}

\wraptikz{%
  \begin{figure}[tp]
    \centering
    \begin{subfigure}[b]{0.495\columnwidth}
      \centering
      \includepdf{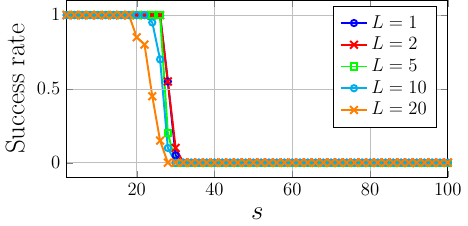}
      \caption{
        Canonical basis with $\dimytot = 500$
      }
    \end{subfigure}
    \begin{subfigure}[b]{0.495\columnwidth}
      \centering
      \includepdf{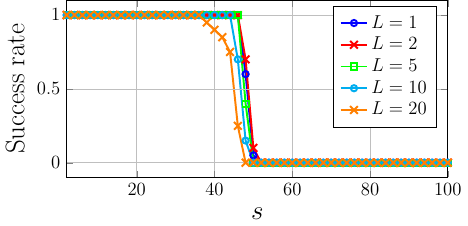}
      \caption{
        Canonical basis with $\dimytot = 740$
      }
    \end{subfigure}
    \par\bigskip
    \begin{subfigure}[b]{0.495\columnwidth}
      \centering
      \includepdf{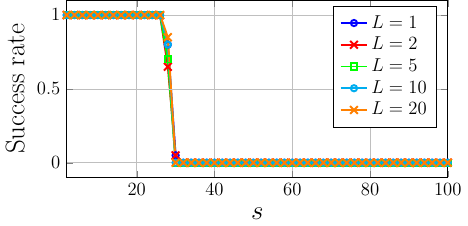}
      \caption{
        \ac{DFT} basis with $\dimytot = 500$
      }
    \end{subfigure}
    \begin{subfigure}[b]{0.495\columnwidth}
      \centering
      \includepdf{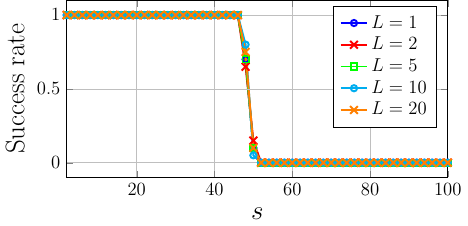}
      \caption{
        \ac{DFT} basis with $\dimytot = 740$
      }
    \end{subfigure}
    \caption{
      Sectional cuts through the phase transition diagrams in
      \blaref{fig:group_phase_transition_canonical} and
      \blaref*{fig:group_phase_transition_fourier} demonstrating the effects of
      varying numbers of sensors on the recovery performance for different
      sparsity bases
    }
    \label{fig:group_phase_transition_sections}
  \end{figure}%
}

{We repeat the same experiment for group-sparse signals in the frequency domain,
\ie, we set $\vmPsi = \vmF_\dimxtot$.
The results are shown in \blaref{fig:group_phase_transition_fourier}.
As predicted by \blaref{thm:block_diagonal_group_rip}, the effects of varying
$\numO$ are even less pronounced than in case of the canonical basis since
neither the previous behavior around $\dimy = 0.1\dimx$, nor the widening of
the transition zone can be observed.
This confirms the intuition that the incoherence of the Fourier basis with
the canonical basis allows for a reduction in the number of measurements per
sensor without affecting the overall reconstruction performance.
To inspect this behavior a little closer, we additionally plot two sections
through each phase transition diagram for $\dimytot = 500$ and $\dimytot =
740$ in \blaref{fig:group_phase_transition_sections}.
This representation clearly demonstrates the diminishing performance with an
increased number of measurements for canonically group-sparse vectors.
For frequency group-sparse vectors, however, the performance is invariant under
the choice of $\numO$.}

\wraptikz{%
  \begin{figure}[tp]
    \centering
    \begin{subfigure}[b]{0.495\columnwidth}
      \centering
      \includepdf{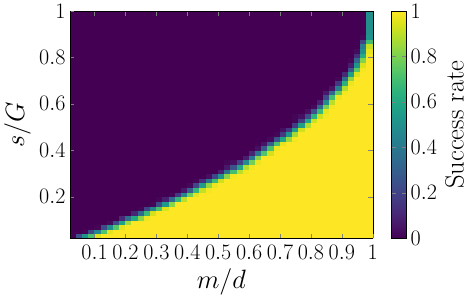}
      \caption{$\numO = 1$, $\vmPsi = \Id_\dimxtot$}
    \end{subfigure}
    \begin{subfigure}[b]{0.495\columnwidth}
      \centering
      \includepdf{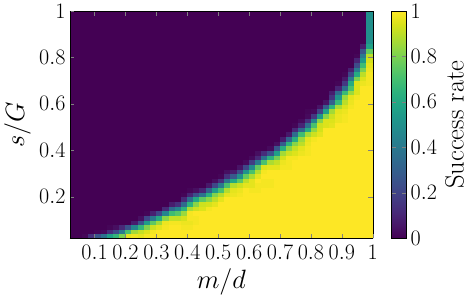}
      \caption{$\numO = 20$, $\vmPsi = \Id_\dimxtot$}
    \end{subfigure}
    \par\bigskip
    \begin{subfigure}[b]{0.495\columnwidth}
      \centering
      \includepdf{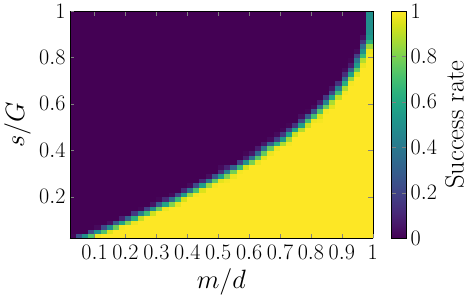}
      \caption{$\numO = 1$, $\vmPsi = \vmF_\dimxtot$}
    \end{subfigure}
    \begin{subfigure}[b]{0.495\columnwidth}
      \centering
      \includepdf{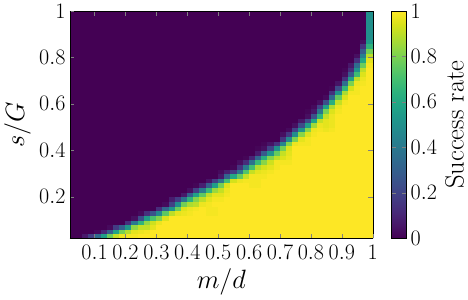}
      \caption{$\numO = 20$, $\vmPsi = \vmF_\dimxtot$}
    \end{subfigure}
    \caption{
      Phase transition diagrams for different numbers of sensors and sparsity
      bases when each sensor is equipped with an identical copy of the
      prototype subgaussian measurement matrix $\vmPhi \in \R^{\dimy \times
      \dimx}$
    }
    \label{fig:group_phase_transition_canonical_and_fourier_rbd}
  \end{figure}%
}

{Finally, we conduct the same experiments as before for the scenario in which
each sensor is equipped with a copy of the same random matrix $\vmPhi \in
\R^{\dimy \times \dimx}$ which is drawn once and then fixed throughout all
subsequent experiments.
The results are shown in
\blaref{fig:group_phase_transition_canonical_and_fourier_rbd}.
The phase transition diagrams confirm the assumption that the general recovery
behavior is comparable to the previous setting given the identical dependence
of $\dimy$ on $\dims,\dimg,\dimxtot$ and $\numG$ predicted by
\blaref{thm:same_block_diagonal_group_rip}.
More precisely, we observe a similar widening of the transition zone as the
number of sensors increases both for the canonical and the Fourier basis, as
well as a reduced tapering of the phase transition diagrams for small $\dimy$.
In contrast to the scenario in which we equip each sensor with an independent
sensing matrix, the sectional cuts through the individual diagrams depicted in
\blaref{fig:group_phase_transition_sections_rbd} further reveal a slight drop
in recovery performance for the \ac{DFT} basis as the number of sensors
$\numO$ increases.
This effect is likely captured by the parameter $\orthoconst(\vmPsi)$,
which---due to its complicated nature---does not admit a straightforward
calculation and interpretation for $\vmPsi = \vmF_\dimxtot$ as the coherence
parameter $\coherence(\vmPsi)$ in the previous setting.
Finding a more meaningful bound for $\orthoconst(\vmPsi)$ therefore remains an
interesting open problem in this context.}

\wraptikz{%
  \begin{figure}[tp]
    \centering
    \begin{subfigure}[b]{0.495\columnwidth}
      \centering
      \includepdf{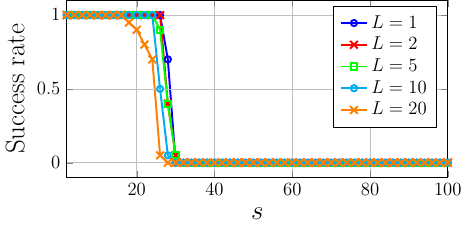}
      \caption{
        Canonical basis with $\dimytot = 500$
      }
    \end{subfigure}
    \begin{subfigure}[b]{0.495\columnwidth}
      \centering
      \includepdf{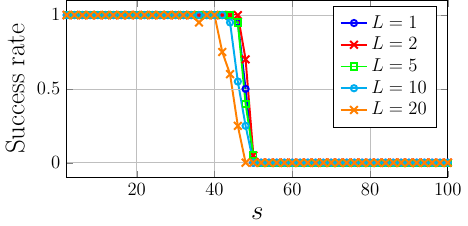}
      \caption{
        Canonical basis with $\dimytot = 740$
      }
    \end{subfigure}
    \par\bigskip
    \begin{subfigure}[b]{0.495\columnwidth}
      \centering
      \includepdf{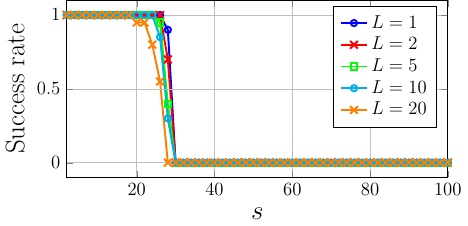}
      \caption{
        \ac{DFT} basis with $\dimytot = 500$
      }
    \end{subfigure}
    \begin{subfigure}[b]{0.495\columnwidth}
      \centering
      \includepdf{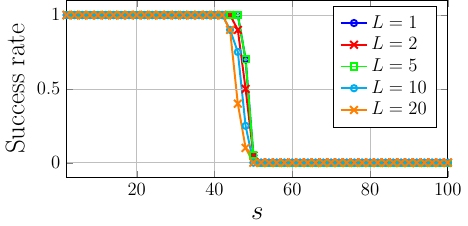}
      \caption{
        \ac{DFT} basis with $\dimytot = 740$
      }
    \end{subfigure}
    \caption{
      Sectional cuts through the phase transition diagrams in
      \blaref{fig:group_phase_transition_canonical_and_fourier_rbd}
      demonstrating the effects of varying numbers of sensors on the recovery
      performance for different sparsity bases
    }
    \label{fig:group_phase_transition_sections_rbd}
  \end{figure}%
}

\section{Conclusion}
\label{sec:conclusion}

In this paper, we established conditions on the number of measurements required
to stably and robustly estimate group-sparse vectors by means of block diagonal
measurement matrices whose blocks either consist of independent \emph{or}
identical copies of subgaussian matrices. Appealing to a powerful concentration
bound on the suprema of chaos processes, we derived conditions on the number of
measurements required for subgaussian block diagonal random matrices to satisfy
the so-called \acl{GRIP}. This generalizes an earlier result due to Eftekhari
\etal who first established a similar result for the canonical sparsity model.
Although certain adversarial group partitions including the distributed sensing
model may lead to suboptimal scaling of the number of measurements, such cases
are generally avoided if signals are group-sparse in nonlocalized sparsity
bases whose basis matrices are not block diagonal. In this case, our results
predict almost optimal scaling behavior up to logarithmic factors. {An interesting future work is to remove some of the logarithmic terms that appear in the bound. Some of these dependencies appear as well in similar works on bounded orthogonal systems, for example, the logarithmic dependency on the ambient dimension $D$ or dependency on higher powers of $\log(s)$. These are conjectured to be removable, which constitute a challenging open problem.}

\section*{Acknowledgment}

The authors would like to thank Holger Rauhut for many fruitful discussions on
the topics addressed in the present paper.

\appendix

\section{Proof of \blaref{thm:group_rip_guarantee}}
\label{app:proof_thm_group_rip}

In general, necessary and sufficient conditions for sparse recovery depend on
the so-called \ac{NSP} which ensures that the null space of the measurement
matrix does not contain any sparse vectors of a certain order besides the zero
vector. In this section, we provide a similar sufficient condition for
group-sparse recovery. The group-sparse \acs{NSP} is a natural generalization
of the block-spare \acs{NSP}, which was originally introduced in
\cite{Gao2017ANB}. Similar to the proofs in the block-sparse case of \opcit,
the structure of our proof follows the example of the respective proof in the
canonical sparsity setting (\cf \cite[\chapName~4 and
6]{Foucart2013mathematicalIC}).

\subsection{Robust \texorpdfstring{\Acs{GNSP}}{Group-NSP}}

\Definition[$\ell_2$-robust \acs{GNSP}]{\label{def:group_nsp}
  Given $q\geq 1$, a matrix $\vmA\in\C^{\dimytot\times \dimxtot}$ is said to
  satisfy the $\ell_2$-robust \ac{GNSP} of order $s$ with respect to
  $\norm{\cdot}$ and constants $\rho\in(0,1)$ and $\tau>0$ if for all
  $\vmv\in\C^\dimxtot$ and for all $S\subset\set{1,\ldots,G}$ with
  $\card{S}=s$,
  \Equation*{
    \norm{\vmv_{\partI_S}}_{2}\leq \frac{\rho}{\sqrt s}
    \norm{\vmv_{\partI_{{\comp{S}}}}}_{\partI,1}+\tau\norm{\vmA\vmv}.
  }
}
{From the inequality
$\norm{\vmv}_{{\partI_S},1}\leq \sqrt{s}\norm{\vmv_{\partI_S}}_{2}$,
it can be seen that $\ell_2$-robust group-NSP implies the following:
  \Equation*{
    \norm{\vmv}_{{\partI_S},1}\leq {\rho}
    \norm{\vmv_{\partI_{{\comp{S}}}}}_{\partI,1}+\tau{\sqrt s}\norm{\vmA\vmv}.
  }
This is the so called $\ell_1$-robust group-NSP condition, which is weaker than $\ell_2$-robust group-NSP.
}

\Theorem{\label{thm:GNSP}
  Suppose that the matrix $\vmA\in\C^{\dimytot\times \dimxtot}$ satisfies the
  $\ell_2$-robust \acl{GNSP} of order $s$ with respect to $\norm{\cdot}$
  and constants $\rho\in(0,1)$ and $\tau>0$. Then for any $\vmx,\vmz\in\C^D$,
  \Equation*{
    \norm{\vmz-\vmx}_2
    &\leq \frac{C}{\sqrt s}\parens{\norm{\vmz}_{\partI,1}-\norm{\vmx}_{\partI,1}
    + 2\sigma_s(\vmx)_{\partI,1}} \\
    &\quad+ D\norm{\vmA(\vmz-\vmx)}
  }
  where $C=\frac{(1+\rho)^2}{1-\rho}$ and $D=\frac{(3+\rho)\tau}{1-\rho}$.
}
\Proof{
  We introduce the following notation used throughout the rest of the proof.
  Given a group partition $\partI = \set{\partI_1, \ldots, \partI_\numG}$ and a
  group index set $S \subset [\numG]$, we denote by $\partI_S$ the subpartition
  $\smallset{\partI_i}[i \in S]$. Moreover, we denote by $\partI_\comp{S}$ the
  subpartition consisting of the groups indexed by $\comp{S} = [\numG]
  \setminus S$. Finally, with slight abuse of notation, we write
  $\vmx_{\partI_S}$ for the vector $\vmx \in \C^\dimxtot$ restricted to the
  index set $\union_{i \in S} \partI_i$.

  The $\ell_2$-robust \acs{GNSP} directly implies that for any
  $\vmx,\vmz\in\C^D$ and $\vmv=\vmz-\vmx$, we have
  \Equation{
    \norm{\vmv}_{2}&\leq
    \norm{\vmv_{\partI_S}}_{2}+\norm{\vmv_{\partI_{\comp{S}}}}_{2}\notag \\
    &\leq \frac{\rho}{\sqrt s}
    \norm{\vmv_{\partI_{\comp{S}}}}_{\partI,1} +
    \tau\norm{\vmA\vmv}+\norm{\vmv_{\partI_{\comp{S}}}}_{2}\label{eqn:NSP1}.
  }
  We first provide a bound for $\norm{\vmv_{\partI_{\comp{S}}}}_{2}$ in terms
  of $\norm{\cdot}_{\partI,1}$. Denote by
  $\set{\partI_1^*,\ldots,\partI_G^*}$ the nonincreasing group rearrangement of
  $\partI$ such that
  \Equation*{
    \norm{\vmv_{\partI_1^*}}_{2}
    \geq \norm{\vmv_{\partI_2^*}}_{2}
    \geq \dots\geq \norm{\vmv_{\partI_G^*}}_{2}.
  }
  We choose $S$ as the index set of the best $s$-term group approximation of
  $\vmv$ which implies that
  \Equation*{
    \norm{\vmv_{\partI_{\comp{S}}}}_{2}^2
    &\leq \sum_{j=s+1}^G \norm{\vmv_{\partI_j^*}}^2_{2} \\
    &\leq \parens{\frac 1s\sum_{j=1}^s
    \norm{\vmv_{\partI_j^*}}_{2}} \parens{\sum_{j=s+1}^G
    \norm{\vmv_{\partI_j^*}}_{2}} \\
    &\leq \frac{1}{s}\norm{\vmv}_{\partI,1}^2.
  }
  Applying this inequality to \blaref*{eqn:NSP1} therefore yields
  \Equation{
    \norm{\vmv}_{2}\leq \frac{1+\rho}{\sqrt s}
    \norm{\vmv}_{\partI,1}+\tau\norm{\vmA\vmv}. \label{eqn:NSP2}
  }
  Next we bound $\norm{\vmv}_{\partI,1}$. First note that if the $\ell_2$-robust
  \acs{GNSP} holds, the Cauchy-Schwarz inequality implies the following bound
  on the group $\ell_1$-norm:
  \Equation{\label{eqn:group_l1_bound}
    \norm{\vmv_{\partI_S}}_{\partI,1}\leq {\rho}
    \norm{\vmv_{\partI_{\comp{S}}}}_{\partI,1}+\tau{\sqrt s}\norm{\vmA\vmv}.
  }
  Invoking \blaref{eqn:group_l1_bound}, we have
  \Equation{
    \norm{\vmv}_{\partI,1}&=\norm{\vmv_{\partI_S}}_{\partI,1}+
    \norm{\vmv_{\partI_{\comp{S}}}}_{\partI,1}\notag\\
    &\leq (1+\rho) \norm{\vmv_{\partI_{\comp{S}}}}_{\partI,1}+\tau{\sqrt s}
    \norm{\vmA\vmv} \label{eqn:NSP3}.
  }
  Here $S$ can be chosen differently as before. We apply
  \blaref{eqn:group_l1_bound} once again in combination with the following
  result which is easily adapted to the group-sparse setting from
  \cite[\lemName~4.15]{Foucart2013mathematicalIC}.

  \Lemma{\label{lem:mixednorminequality}
    Consider group-sparse signals with $G$ groups and partition $\partI$. For
    $S\subset[G]$, vectors $\vmx,\vmz\in\C^D$ and $\vmv=\vmz-\vmx$, we have
    \Equation*{
      \norm{\vmv_{\partI_{\comp{S}}}}_{\partI,1} \leq
      \norm{\vmz}_{\partI,1}-\norm{\vmx}_{\partI,1}+
      \norm{\vmv_{\partI_{S}}}_{\partI,1}+2\norm{\vmx_{\partI_{\comp{S}}}}_{\partI,1}.
    }
  }

  We apply \blaref{eqn:group_l1_bound} to the above inequality to obtain
  \Equation*{
    \norm{\vmv_{\partI_{\comp{S}}}}_{\partI,1} &\leq
    \norm{\vmz}_{\partI,1}-\norm{\vmx}_{\partI,1}+
    \norm{\vmv_{\partI_{S}}}_{\partI,1}+2\norm{\vmx_{\partI_{\comp{S}}}}_{\partI,1}\\
    &\leq \norm{\vmz}_{\partI,1}-\norm{\vmx}_{\partI,1}+ {\rho}
    \norm{\vmv_{\partI_{\comp{S}}}}_{\partI,1}\\
    &\quad +\tau{\sqrt
    s}\norm{\vmA\vmv}+2\norm{\vmx_{\partI_{\comp{S}}}}_{\partI,1},
    }
    which implies that
    \Equation*{
      \norm{\vmv_{\partI_{\comp{S}}}}_{\partI,1} &\leq
      \frac{1}{1-\rho}\parens{\norm{\vmz}_{\partI,1}-\norm{\vmx}_{\partI,1}+
      2\norm{\vmx_{\partI_{\comp{S}}}}_{\partI,1} } \\
      &\quad+\frac{\tau{\sqrt s}}{1-\rho}\norm{\vmA\vmv},
    }
    and consequently from \blaref*{eqn:NSP3} that
    \Equation*{
      \norm{\vmv}_{\partI,1} &\leq
      \frac{1+\rho}{1-\rho}\parens{\norm{\vmz}_{\partI,1}-\norm{\vmx}_{\partI,1}+
      2\norm{\vmx_{\partI_{\comp{S}}}}_{\partI,1}} \\
      &\quad +\frac{2\tau{\sqrt s}}{1-\rho}\norm{\vmA\vmv}.
    }
  The choice of $S$ to minimize the right hand side is the support for the best
  $s$-term group approximation of $\vmx$. Combined with \blaref*{eqn:NSP2}, the
  theorem follows.
}

Since the result above holds for any $\vmx$ and $\vmz$, choosing $\vmx=\xorg$
and $\vmz=\opt{\vmx}$ with $\opt{\vmx}$ denoting a minimizer of
\blaref{prob:group_qcbp} immediately implies the following theorem.

\Theorem{\label{thm:NSP_group_qcbp}
  Suppose that ${\vmA} \in \C^{\dimytot \times \dimxtot}$ satisfies the
  $\ell_2$-robust \acs{GNSP} of order $s$ with constants $0<\rho<1$ and
  $\tau>0$. Then for all $\xorg \in \C^\dimxtot$, and $\vmy = \tilde{\vmA}\xorg +
  \vme$ with $\norm{\vme}_2 \leq \epsilon$, any solution $\opt{\vmx}$ of the
  program \blaref{prob:group_qcbp} approximates $\vmx$ with error
  \Equation*{
    \norm{\xorg - \opt{\vmx}}_2 \leq {C}
    \frac{\ktermerror{\dims}(\xorg)_{\partI,1}}{\sqrt{\dims}} + {D}
    \epsilon
  }
  with $C,D > 0$.
}

The $\ell_2$-robust \acs{GNSP} provides a necessary and sufficient
condition for recovery of group-sparse vectors.
In the next section, we establish that the \ac{GRIP} implies the robust
\ac{GNSP} and therefore yields a sufficient condition for stable and robust
recovery of group-sparse vectors.
{As we have seen above, the $\ell_2$-robust group-NSP implies the $\ell_1$-robust group-NSP. Replicating, the above proof for $\ell_1$-robust group-NSP, Theorem \ref{thm:NSP_group_qcbp} implies:
  \Equation*{
    \norm{\xorg - \opt{\vmx}}_2 \leq {C}
  {\ktermerror{\dims}(\xorg)_{\partI,1}} + {D}{\sqrt{\dims}}
    \epsilon
  }
}

\subsection{\texorpdfstring{\Acs{GRIP}}{Group-RIP} and Robust \texorpdfstring{\Acs{GNSP}}{Group-NSP}}

In light of the previous section, it suffices to prove that the group-\acs{RIP}
of order $2s$ with constant $\delta$ implies the robust \acs{GNSP} in
order to prove \blaref{thm:group_rip_guarantee}. Inspired by
\cite[\chapName~6]{Foucart2013mathematicalIC}, consider the sets
$S_0,S_1,\dots$ such that $S_i$ is defined as the index of $s$ largest groups
in $\comp{\union_{j<i} S_j}$. If the \acs{GNSP} is established for
$S_0$, which yields the largest possible $\norm{\vmv_{\partI_{S_0} }}_2$, then
it holds also for all $S$. Assuming the \acs{GRIP} holds, we have
$\norm{\vmA\vmv_{\partI_{S_0} }}_2^2=(1+t)\norm{\vmv_{\partI_{S_0} }}_2^2$ with
$\abs{t}<\delta$ and therefore, we can bound $\norm{\vmA\vmv_{\partI_{S_0}
}}_2^2$ by
\Equation*{
\norm{\vmA\vmv_{\partI_{S_0} }}_2^2&=\inner{{\vmA\vmv_{\partI_{S_0} }}} {\vmA(\vmv-\sum_{k\geq 1}\vmv_{ \partI_{S_k} })} \\
&=\inner{{\vmA\vmv_{\partI_{S_0} }}} {\vmA\vmv}-\sum_{k\geq 1}\inner{{\vmA\vmv_{\partI_{S_0} }}} {\vmA\vmv_{ \partI_{S_k} }}\\
&\leq \norm{{\vmA\vmv_{\partI_{S_0} }}}_2\norm{\vmA\vmv}_2+C_t\sum_{k\geq 1}\norm{{\vmv_{\partI_{S_0} }}}_2 \norm{\vmv_{ \partI_{S_k} }}_2
}
where the last inequality follows from \blaref{lem:NSP-RIP1} given at the end
of this section with $C_t=\sqrt{\delta^2-t^2}$. Using
$\norm{{\vmA\vmv_{\partI_{S_0} }}}_2=\sqrt{1+t}\norm{\vmv_{\partI_{S_0} }}_2$ we
arrive at an expression similar to the \acs{GNSP}, namely:
\Equation{\label{eqn:GNSP_RIP}
 (1+t)\norm{\vmv_{\partI_{S_0} }}_2\leq C_t\sum_{k\geq 1}\norm{\vmv_{ \partI_{S_k} }}_2+\sqrt{1+t}\norm{\vmA\vmv}_2.
}
Although $\sum_{k\geq 1}\norm{\vmv_{ \partI_{S_k} }}_2\leq \norm{\vmv_{ \partI_{
\comp{S_0} } }}_{\partI,1}$, we need an additional $1/\sqrt s$ term to get the
\acs{GNSP}. Invoking \cite[\lemName~6.14]{Foucart2013mathematicalIC}, we
immediately obtain
\Equation*{
  \sum_{k\geq 1}\norm{\vmv_{ \partI_{S_k} }}_2\leq \frac{1}{\sqrt s}\norm{\vmv_{
    \partI_{ \comp{S_0} } }}_{\partI,1}+\frac 14 \norm{\vmv_{ \partI_{S_0} }}_2.
}
Next, we apply the above inequality to \blaref*{eqn:GNSP_RIP} which---after
standard manipulations---yields
\Equation*{
  \norm{\vmv_{\partI_{S_0} }}_2
  &\leq \frac{\delta}{\sqrt{1-\delta^2}-\delta^2/4}\frac{1}{\sqrt
  s}\norm{\vmv_{ \partI_{ \comp{S_0} } }}_{\partI,1} \\
  &\quad
  +\frac{\sqrt{1+\delta}}{\sqrt{1-\delta^2}-\delta^2/4}\norm{\vmA\vmv}_2.
}
Therefore the \acs{GNSP} holds with $\rho$ and $\tau$ given by
\Equation*{
\rho=\frac{\delta}{\sqrt{1-\delta^2}-\delta^2/4}\text{  and  } \tau=\frac{\sqrt{1+\delta}}{\sqrt{1-\delta^2}-\delta^2/4}.
}
This holds provided that $\rho<1$ which is equivalent to $\delta<4/\sqrt{41}$.
The constants $C$ and $D$ follow accordingly. The claim follows.

It remains to establish the following result.

\Lemma{\label{lem:NSP-RIP1}
  Suppose that the matrix $\vmA\in\C^{\dimytot\times\dimxtot}$ satisfies the
  group-\acs{RIP} of order $2s$. For two disjoint sets $S_0,S_1\subset [G]$
  with cardinality $s$,
  \Equation*{
    \abs{\inner{\vmA\vmv_{\partI_{S_0} }}{\vmA\vmv_{\partI_{S_1} }}}\leq
    \sqrt{\delta^2-t^2}\norm{\vmv_{\partI_{S_0}}}_2\norm{\vmv_{\partI_{S_1}}}_2.
  }
}
\Proof{
To start with, we normalize the two vectors to have unit $\ell_2$-norm by
defining the auxiliary vectors $\vmu := \vmv_{\partI_{S_0}} /
\norm{\vmv_{\partI_{S_0}}}_2$ and $\vmw := \vmv_{\partI_{S_1}} /
\norm{\vmv_{\partI_{S_1}}}_2$.
Fix $\alpha,\beta>0$. Then
  \Equation*{
    2\abs{\inner{\vmA\vmu}{\vmA\vmw}}&=\frac{1}{\alpha+\beta}(\norm{\vmA(\alpha\vmu+\vmw)}_2^2-\norm{\vmA(\beta\vmu-\vmw)}_2^2\\
    &-(\alpha^2-\beta^2)\norm{\vmA\vmu}_2^2)\\
    &\leq \frac{1}{\alpha+\beta}[(1+\delta)\norm{\alpha\vmu+\vmw}_2^2\\
    &-(1-\delta)\norm{\beta\vmu-\vmw}_2^2-(\alpha^2-\beta^2)(1+t)\norm{\vmu}_2^2]\\
    &\leq \frac{1}{\alpha+\beta}[(1+\delta)(\alpha^2+1)^2\\
    &-(1-\delta)({\beta^2+1})^2-(\alpha^2-\beta^2)(1+t)]\\
    &\leq \frac{1}{\alpha+\beta}[\alpha^2(\delta-1)+\beta^2(\delta+1)+2\delta].
  }
  Choosing $\alpha=(\delta+t)/\sqrt{\delta^2-t^2}$ and
  $\beta=(\delta-t)/\sqrt{\delta^2-t^2}$ completes the proof.
}

\printbibliography

\end{document}